\documentclass[]{article}
\usepackage{multicol}
\usepackage{apj}

\newcommand {\ks} {km~s$^{-1} \;$}

\def\lesssim{\mathrel{\hbox{\rlap{\hbox{\lower4pt\hbox{$\sim$}}}\hbox{$<$}}}}
\def\gtrsim{\mathrel{\hbox{\rlap{\hbox{\lower4pt\hbox{$\sim$}}}\hbox{$>$}}}}

\begin{document}

\vspace{15mm}
\begin{center}
\uppercase{GALAXY DISTANCES IN THE NEARBY UNIVERSE: \\
CORRECTIONS FOR PECULIAR MOTIONS}\\   
\vspace*{1.5ex}
{\sc Christian MARINONI$^{1,2}$, Pierluigi MONACO$^{3}$, 
Giuliano GIURICIN$^{1,2}$, and Barbara COSTANTINI$^{1,2}$}\\
\vspace*{1.ex}
{\small $^1$ Dipartimento di Astronomia,
Universit\`{a} di Trieste, via Tiepolo 11, 34131 Trieste, Italy\\
$^2$ SISSA, via Beirut 4, 34013 - Trieste, Italy, \\
$^3$ Institute of Astronomy, Madingley Road, Cambridge, CB3 OHA, UK\\
e-mail: marinoni@mizar.sissa.it;
monaco@ast.cam.ac.uk;
giuricin@newton.sissa.it;\\ 
barbara@newton.sissa.it}
\end{center}
\vspace*{-6pt}

\begin{abstract}

By  correcting the redshift--dependent  distances for peculiar motions
through  some peculiar  velocity field   models,  we recover the  true
distances of an  extensive,  all-sky sample of nearby  galaxies ($\sim
6400$ galaxies with  recession  velocities  $cz<5500\;$\ks),  which is
complete    up   to   the     apparent     limiting blue     magnitude
$B=14\;$mag. Relying on available catalogs  of galaxy groups, we treat
$\sim$2700 objects  as  members  of galaxy  groups and  the  remaining
objects as field galaxies.

We  invert  the  derived    redshift-distance relations   to  estimate
distances for field galaxies and  groups and we overcome the ambiguity
inherent  to the triple-valued   zones by using Tully-Fisher relations
calibrated on  suitably defined samples  of galaxies  having distances
predicted by peculiar velocity models.

We use two  independent approaches to  modeling the  peculiar velocity
field: i)  a cluster dipole  reconstruction scheme that we modify with
the    inclusion of a local      model of Virgocentric-infall; ii)   a
multi--attractor model fitted to the Mark II  and Mark III catalogs of
galaxy peculiar  velocities.  In the multi--attractor model  we assume
that the velocity field is generated  by a few prominent gravitational
sources      (Virgo     cluster, "Great   Attractor",   Perseus-Pisces
Supercluster, and Shapley supercluster).

We discuss differences in   the  results  coming out from    different
velocity  models  and  from  different   Mark  II and  Mark III   data
subsets. In particular, according to Mark III data the Great Attractor
appears to have a smaller  influence on local dynamics than previously
believed, whereas the Perseus-Pisces and Shapley superclusters acquire
a specific dynamical role. Remarkably, the Shapley structure, which is
found  to  account for  nearly half  the peculiar  motion of the Local
Group, is placed by Mark III data closer to the zone of avoidance with
respect  to  its optical  position.  On  the other  hand, the modified
cluster   dipole model is   characterized  by  relatively small  flows
towards the  Great  Attractor and the   Shapley supercluster, together
with a large Virgocentric infall.

Our multi--attractor model  based    on   Mark  III data   favors    a
cosmological  density  parameter $\Omega_0\sim$0.5  (irrespective of a
biasing factor of order unity).
 
The use of different peculiar velocity field models allows us to check
to what extent differences in current views on cosmic flows affect the
recovering  of galaxy   distances.   We find  that  differences  among
distance estimates are less  pronounced in the  $\sim 2000 - 4000$ \ks
distance  range than at  larger  or  smaller  distances.  In the  last
regions these differences have a serious impact  on the 3D maps of the
galaxy distribution  and on  the local  galaxy  density  ---  on small
scales ($<$  1 Mpc) ---, which  is  a crucial parameter  being used in
statistical   studies of environmental  effects   on the properties of
nearby galaxies.

\vspace*{6pt}
\noindent
{\em Subject headings: }
galaxies: distances and redshifts --- 
galaxies: clusters: general --- cosmology: large--scale structure of 
universe

\end{abstract} 
\begin{multicols}{2}

\section{INTRODUCTION} 

The determination of  galaxy distances is one  of the most fundamental
problem in  extragalactic astronomy.  In particular, galaxy  distances
allow  us to map the  three--dimensional (3D) distribution of galaxies
and hence to evaluate the local galaxy  density, which is an important
characteristics of the galaxy environment.

In many previous studies which rely only on the two-dimensional (2D)
projection of galaxies on the sky (see, e.g., Lahav (1987) for all-sky
2D optical maps of bright galaxies), a local galaxy density parameter is 
defined in terms of galaxy counts over a sky area which are statistically
corrected for projection effects. With the advent of large surveys of
galaxy redshifts and well-selected galaxy catalogs, it has become possible
to map the spatial distribution of galaxies and hence to construct a
3D definition of the local galaxy density. 

Optical galaxy samples are more suitable for mapping the galaxy density
field on relatively small scales than IRAS-selected galaxy samples, which
have been frequently used as tracers of the density field on large scales
up to a recession velocity $cz\sim20000\;$ \ks (e.g., Rowan-Robinson et
al., 1990; Kaiser et al., 1991; Strauss et al., 1992a,b; Fisher et al.,
1995). The latter samples have the advantage that they are homogeneously
selected and penetrate to lower galactic latitudes because IRAS fluxes are
much less impeded by Galactic extinction. But the IRAS galaxy samples have
the serious drawback that they do not include the early-type galaxies
(because of their little dust content and star formation), which usually
lie in the densest regions of clusters and superclusters.  Furthermore,
the density field traced by IRAS galaxies is noisy nearby, because of the
sparseness of the samples and because IRAS fluxes are much less linked
with galaxy mass than optical fluxes. Therefore, in the following we shall
focus on all-sky optical galaxy samples. 

The magnitude-limited Revised Shapley-Ames (RSA) Catalog of Bright
Galaxies (Sandage \& Tamman, 1981), which contains the first large
compilation of redshift data for bright galaxies across the entire sky,
was used by Yahil, Sandage \& Tamman (1980) to delineate the 3D-density
field of galaxies in the Local Supercluster (LS).  The structures of the
LS region have been comprehensively described and named by Tully \& Fisher
(1987). The former author intended to include all nearby galaxies with
systemic recession velocities $cz<3000\;$\ks (2367 objects) in his Nearby
Galaxies Catalog (Tully, 1988a, hereafter NBG), which is a combination of
the RSA catalog and a diameter-limited sample of late-type and fainter
galaxies found in an all-sky HI survey made principally by Fisher \& Tully
(1981) and Reif et al. (1982). In the NBG catalog, which is complete up to
the corrected blue total magnitude $B_T\sim12$~mag (although it extends to
fainter magnitudes), the distances of all non-cluster galaxies have been
estimated on the basis of their redshifts, with an assumed Hubble constant
$H_0=75\;km s^{-1} Mpc^ {-1}$, whereas the galaxy members of systems with
relatively high velocity dispersion have been given a distance consistent
with the mean redshift of the system. The important fact that redshifts
are not equivalent to distances because galaxies have peculiar motions
(i.e. deviations from the pure Hubble flow) is simply taken into account
in the NBG catalog through the Virgocentric retardation model described by
Tully \& Shaya (1984), in which the authors assume that the Milky Way is
retarded by 300 \ks from the pure Hubble flow by the mass of the Virgo
cluster. Defining the local galaxy density on the basis of his NBG
catalog, Tully (1988b) incorporated corrections for the catalog
incompletion at large distances. 

Later, similar quantifications of the local galaxy density, based on the
NBG catalog data, have been exploited in statistical analyses of
environmental effects on some properties of LS galaxies such as the
frequency of bars (Giuricin et al., 1993), the classification of spiral
arms (Giuricin et al., 1994), the frequency of active galactic nuclei
(LINER and Seyfert objects) selected from optical spectroscopic surveys
(Monaco et al., 1994), and the bulge-to-disk light ratio (Giuricin et al.,
1995). 

In an effort of going beyond the LS, Hudson (1993a,b; 1994a,b)
assembled an extensive galaxy sample from a merging of the
diameter-limited northern UGC catalog (Nilson, 1973) and the diameter-
limited southern ESO catalog (Lauberts, 1982; Lauberts \& Valentijn,
1989). He took into account the fairly large incompleteness in redshift of
his sample as a function of angular diameter and position on the sky, by
applying statistical corrections, which allowed him to reconstruct the
density field of optical galaxies to a depth of $cz=8000$ \ks. 

The Lyon-Meudon Extragalactic Database (LEDA), which collects and
homogenizes several data for all the galaxies of the main optical
catalogues, such as the catalogs UGC (Nilson, 1973), ESO (Lauberts, 1982,
Lauberts \& Valentijn, 1989), CGCG (Zwicky et al., 1961-1968), and ESGC
(Corwin \& Skiff, 1998, in preparation), allows the extraction of an
all-sky galaxy sample which has properties of completeness and is deep
enough to probe the galaxy field significantly beyond the LS. In
particular, Garcia et al. (1993) extracted a magnitude-limited sample
which covers all the sky and contains 6392 galaxies having recession
velocities $cz<5500$ \ks. Although different optical catalogs are
characterized by different limits of completeness in apparent magnitude or
angular diameter, the above-mentioned galaxy sample was found to be
substantially complete up to its limiting corrected total blue magnitude
$B_T=14$~mag (for galactic latitudes $|b|>20^{\circ}$) (Garcia et al.,
1993).  The authors tabulated several parameters for each galaxy, such as
the morphological type, the maximum velocity rotation deduced from the
deprojected 21-cm hydrogen line width, the de Vaucouleurs' luminosity
index, the distance modulus, the corrected angular sizes and total blue
magnitudes. The authors derived the last two quantities by transforming
the original raw data to the standard system of the RC3 catalog (de
Vaucouleurs et al., 1991) and by applying corrections for Galactic
extinction, internal extinction, and K-dimming. Only for the few galaxies
which lack magnitude measures, the values of $B_T$ are rough estimates
derived from their diameters through a mean magnitude-diameter relation. 
The tabulated distance modulus was a weighted mean derived from the
combination of redshift-distances (with $H_0=75\;km\;s^{-1}
Mpc^{-1}$) and redshift-independent distances obtained from distance
indicators (DIs), i.e. from a Tully-Fisher (TF) relation (which relates
the absolute magnitude $M_B$ to the logarithmic maximum rotation velocity
$\log V_m$) and a luminosity-luminosity index relation.  In particular,
the authors used the former kind of distances alone for the farthest
galaxies (with $cz>1500\;$ \ks) and the latter kind alone for the few
nearest objects (with $cz<500\;$ \ks). 

Group assignments for the galaxies of this sample have been provided by
Garcia (1983). She employed two 3D methods of group identification, the
percolation or {\it friends of friends} method proposed by Huchra \&
Geller (1982) and the hierarchical clustering method (Materne, 1978;
Tully, 1980, 1987). The latter method gave, on average, smaller groups
than the former. The adopted final catalog of groups was defined as that
one which includes only groups (as well as galaxy members) common to the
two catalogs.  Accordingly, 3381 objects are field galaxies. Of the
remaining grouped objects, 2703 galaxies are members of 485 systems with
at least three members; most of them are groups which contain less than 10
members, and 13 systems have more than 20 members (among which the clusters
Virgo, Centaurus, Perseus). 

A somewhat different approach to the construction of a large and deep
sample of optical galaxies with good completeness in redshift was followed
by Santiago et al. (1995). Their "Optical Redshift Survey" is a redshift
survey of an optical galaxy sample which contains 8266 galaxies with known
redshift. This sample consists of two largely overlapping subsamples
(which are limited in magnitude and in diameter, respectively) drawn from
the catalogs UGC, ESO, ESGC. The authors selected their own sample
according to the raw (observed) magnitudes and diameters (rather than the
corresponding corrected values) and then quantified the effects of
Galactic extinction (as well as random and systematic errors) on the
optical density field. Redshift-distances were used in general, but the
members of rich clusters were placed at each cluster's center. Probing the
density field at large depth (up to $cz=8000$~ \ks) requires much care in
the derivation of appropriate selection functions, which quantify the loss
of galaxies due to magnitude or diameter limits or other observational
selection effects. Taking into account possible non-uniformities in galaxy
sampling, the authors attempted to match the original three catalogs from
which their sample was drawn, by using separate selection functions
(Santiago et al., 1996). 

The reliability of the distances given in all above-mentioned galaxy
samples (and hence of the resulting galaxy density field) is weakened by
the fact that simple redshift--distances are in general adopted with no
corrections (or no adequate corrections) for the peculiar motions which
are due to inhomogeneous mass distribution and which prevent redshift to
scale linearly with distance. 

In the present paper we propose to recover the true distances of a sample
of nearby galaxies by treating adequately the effects of peculiar
velocities of galaxies. As shown in a forthcoming paper
 (Marinoni et al., 1998b), in
which the resulting 3D galaxy density field is presented and local galaxy
density parameters (corresponding to various smoothing scales) are
discussed, corrections of galaxy distances for peculiar motions appear to
have a large impact on the values of the aforementioned parameters on
small scales (i.e. on $<1$ Mpc scale) and, hence, on the characterization 
of the environment (see \S 6).  

For our purpose, we take the above-mentioned complete sample
extracted from LEDA (Garcia et al., 1993; Garcia, 1993), for which all
galaxy data as well as group assignments are available in the literature.
Thus, we limit ourselves to the nearby universe (within $cz=5500$~\ks),  
where the peculiar velocity field is best known and no large
corrections for galaxy sample incompletion are required. However, the 
galaxy sample considered is deep enough to cover interesting regions of 
prominent overdensities, such as the "Great Attractor" (GA) region and the
Perseus-Pisces supercluster.  

The study of the peculiar velocity field resulting from a proper analysis
of large redshift--distance samples (i.e., samples of galaxies having both
redshift and redshift-independent distance estimates) has acquired a
well-established role in the cosmological context (e.g., for testing the
gravitational instability paradigm for the origin of large--scale
structure, for deducing the relative distributions of luminous and dark
matter, and for constraining the value of the cosmological density
parameter $\Omega_0$; see, e.g., the reviews by Dekel, 1994, and Strauss
\& Willick, 1995). 

In general, peculiar velocity analyses proceed along two main lines of
research. One can turn a distribution of galaxies or galaxy systems with
known redshifts into a smoothed mass-density field (under the linear
approximation of the gravitational instability theory) and then predict
from it the peculiar velocity field , e.g., by means of the "dipole
analysis". Despite the complexity of these studies, the velocity vector of
the Local Group (LG) with respect to the reference frame defined by the
cosmic microwave background (CMB) ($v_{LG}\sim 630$ \ks towards $l \sim
277^{\circ}$ and $b \sim 30^{\circ}$ (Lubin \& Villela, 1986; Kogut et
al., 1993)) has been approximately reconstructed in terms of the
gravitational acceleration exerted by various populations of extragalactic
objects (e.g., the references cited in Kolokotronis et al., 1996). 
 
Alternatively, one can face the problem of deriving the true matter
distribution from the observed peculiar velocity field, e.g using
few-parameter toy models or the non-parametric POTENT method developed by
Bertschinger \& Dekel (1989). 

In this paper, following the second kind of approach, we attempt to
correct the Hubble flow for peculiar motions by means of two basic
independent approaches to modeling the velocity field in the nearby
universe:  i) a modified cluster dipole model, which is Branchini \&
Plionis' (1996, hereafter BP96) optical cluster dipole reconstruction
scheme that we modify by including a local model of the Virgocentric
infall in the LS region. We regard the cluster dipole model as a typical
approach in which the peculiar velocity field is self-consistently derived
through iteration techniques from the redshifts and positions of a sample
of objects. ii) a multi-attractor model, which we take as an example of a
parametric toy model which is fitted to catalogs of peculiar velocities.
In the multi-attractor model we assume that the velocity field is
generated by a few prominent gravitational sources (Virgo cluster , Great
Attractor, Perseus-Pisces Supercluster, Shapley concentration). 

We construct these models relying on the various data subsets
contained in the Mark II and Mark III catalogs of peculiar velocities of
galaxies. In particular, the Mark III catalog (Willick et al., 1997a),
which is the most comprehensive homogenized catalog of peculiar velocities
available today, is a merging of several data sets of spiral and
elliptical galaxies.  

Through these peculiar velocity field models we provide
homogeneous estimates of distances for our galaxy sample. Moreover, 
we can evaluate to what extent differences in current views on 
the peculiar motions affect the recovering of galaxy distances 
and, hence, the 3D maps of galaxy distribution.

The outline of our paper is as follows. In \S 2 we describe the properties
of the Mark II and Mark III catalogs which are relevant to our velocity
field models. In \S 3 we present our two basic approaches to modeling the
peculiar velocity field and we discuss the results relative to different
data sets. In \S 4 we address the inversion of the redshift-distance
relation and, in particular, the ambiguity inherent to the triple-valued
zones of this relation. We attempt to solve this problem using blue TF
relations calibrated on suitably defined samples of galaxies having
distances predicted by velocity field models.  In \S 5, outlining some
basic differences among the velocity field models, we discuss the
resulting sets of galaxy distances. In our conclusions (see \S 6) we
mention some developments of peculiar velocity studies and focus on the
effects of our galaxy distance corrections on the determination of the 3D
galaxy density field. 

Throughout, the Hubble constant is $75\;h\;km\;s^{-1}\;Mpc^{-1}$.   
In general, distances are expressed in velocity units (\ks).

\section{The Mark II and Mark III peculiar velocity catalogs} 

Our modeling of the peculiar velocity field is substantially based on
several data subsets contained in the Mark III catalog of radial peculiar
velocities (Willick et al., 1995, 1996, 1997a).  This is a homogenized 
database of redshift-independent distance estimates obtained using the
revised Faber-Jacson ($D_n-\sigma$) relation for early-type galaxies and
both the forward and inverse TF relations for spirals.  Input data come
from different samples of galaxies, but the inferred galaxy distances are
the result of an accurate analysis which carefully takes into account
differences in the observational selection criteria, in the methods of
measurement, in the TF calibration techniques, in statistical bias
effects. The uniformity of corrected data is ensured through
transformations onto a common system and through a mutually consistent TF
calibration for the samples of spirals. The $D_n-\sigma$- inferred
distances for early-type galaxies are rescaled in order to match the
distance scale of spirals. 
  
This catalog is the current evolution of the previous Mark II catalog
compiled by Burstein (1989), which is a merged set of 1184 galaxies
grouped into 704 objects (single galaxies, groups, and clusters). Of these
galaxies, the ellipticals (E) and lenticulars (S0) come from a
combinations of the dataset used by Lyndell-Bell et al. (1988) in their
analysis of the GA region and those by other authors (Lucey \& Carter,
1988; Faber et al., 1989; Dressler \& Faber, 1990) The spiral sample
consists of the Aaronson "good" and "fair" field spirals (see Faber \&
Burstein, 1988, and Aaronson et al., 1989, for the definitions), the
Aaronson cluster spirals, the spirals of de Vaucouleurs \& Peters (1984).
In some cases we present results derived from the Mark II data alone in
order to emphasize the differences between Mark II and Mark III data. 

With nearly 3400 galaxies, the Mark III catalog, which includes Mark II
data with some improvements, is almost three times bigger than its
predecessor and has a better space coverage. It contains new samples of
spirals which cover sky regions poorly mapped in Mark II. Maps of the
spatial distribution of the various subsamples, which probe different
regions of the sky, are presented by Kollatt et al. (1996). In our
analysis we use the following different subsamples, avoiding the
possibility that the same object be included in different samples: 
\begin{enumerate} 

\item the E/S0 galaxy sample (Lynden-Bell et al. 1988;  Faber
et al. 1989; Lucey \& Carter, 1988; Dressler \& Faber, 1990).  It
includes 544 galaxies grouped into 249 objects which map the GA region. 

\item the HMCL cluster sample of spirals (Han \& Mould, 1992) with
36 objects.  

\item the WCF group sample of spirals (Willick, 1991; Courteau, 1992; 
Courteau et al., 1993; Courteau, 1996) which contains 65 groups of galaxies 
covering the Perseus-Pisces region.

\item the MAT data set of spirals (Mathewson, Ford \& Buchorn, 
1992) with 277 groups selected in the southern hemisphere.  

\item the A82 spiral sample (Aaronson et al., 1982) as revised by Tormen
\& Burstein (1995). Of the 359 galaxies contained in the sample, 222 have
been grouped in 67 groups (Willick et al., 1996) and 137 are field
galaxies. 

\end{enumerate}

The analysis of different data subsets allows us to test the stability of
our final results against the particular subsample used. The whole data
set, in which we take together spirals and ellipticals (which should trace
the same velocity field) will help us to give more stringent constraints
on our models than smaller subsamples would. 
 
The accuracy of the peculiar velocities achieved in Mark III catalog is
not only due to a careful homogenization of different data subsets and DIs
for getting a uniform distance scale; it is also due to the new approach
with which the authors treat sample selection (or calibration) effects,
which typically enter in the calibration of the DI being used in
flux-limited samples, and the inferred-distance effects, also called
Malmquist bias, which typically enter when a calibrated DI is used to
infer distances and hence peculiar velocities (e.g. Willick, 1994).
Selection effects are due to the fact that a magnitude limit in the
selection of the sample used for calibration at a fixed {\it true}
distance (e.g., in a cluster) tilts the direct TF regression line ($M_B$
versus $\log V_m$) towards bright $M_B$ at small values of $\log V_m$. 
The homogeneous part of the Malmquist bias arises from the space geometry
in the sense that the inferred distance $d$ underestimates the true
distance $r$ because it is more likely to have been scattered by errors
from $r>d$ rather than from $r<d$, the volume being $\propto r^{2}$.  The
inhomogeneous part of the Malmquist bias arises from number density
fluctuations and tends to enhance systematically the inferred density 
perturbations. 

Clearly, removing the overall Malmquist effect (known as
inhomogeneous Malmquist bias), arising from both volume effects and
density variations, is of fundamental importance to yield unbiased
peculiar velocities. In the Mark III catalog this was done using smoothed
density fields obtained from the IRAS 1.2 Jy redshift survey (Fisher et
al., 1995), with the effect of peculiar velocities corrected for using
linear theory (Yahil et al., 1991).  The resulting field is expected to
represent the general spiral density field on large scales. 

Thus, besides the uncertainties inherent to DI calibration procedures, the
Mark III distances rely also on a model-dependent reconstruction
of the general density field from redshift data. Moreover, redshift limits
in a sample modify the nature of the Malmquist bias corrections applied. 
It would be worse to neglect any correction for the Malmquist
bias; but we have to take into account the possibility that further errors
and systematic effects can affect the data.  Therefore, in our analysis we
increase the tabulated errors on distances by a factor of 3, according to D.
Burstein's advices for the Mark II catalog. 

For all the single objects we have taken the Malmquist-corrected forward
TF distances, since this is the best approach to follow in our statistical
analysis. Our approach belongs to the method I of the "method matrix" of
peculiar velocity analysis illustrated in \S 5.2 of Strauss \& Willick
(1995). On the other hand, the forward TF distances of the spiral groups
are fully corrected only for selection bias, because they are formed using
redshift-space criteria and in this case it is selection bias rather than
Malmquist bias which pertains.  Residual Malmquist bias due to clustered
structures still affects the data of spiral groups. Nevertheless, the
grouping procedure reduce distance uncertainties by a factor $N^{1/2}$, if
$N$ is the number of grouped galaxies. 

\section{Correcting Galaxy Distances through Velocity Field Models} 

At the low redshifts discussed in this paper the Hubble law, $cz=H_{0}r$,
is an excellent approximation in describing the Friedmann expansion of the
universe.  However, galaxies have motions above and beyond their Hubble
recession velocities. Therefore, by combining the cosmological component
of redshift with the peculiar one, the Hubble relation is modified to: 

\begin{eqnarray} \label{Hubble} \lefteqn{ 
cz \approx H_{0}r+[{\bf v(r)}-{\bf v(0)}]{\bf \cdot\hat{r}}+
[(\frac{
{\bf v(r)\cdot\hat{r}-v(0)\cdot\hat{r}}
}
{c}+ {}} \nonumber  \\
                           & &
{}- \frac{({\bf v(0)\cdot\hat{r})(v(r)
\cdot\hat{r}})}{c^{2}})H_{0}r-\frac{({\bf v(r)\cdot\hat{r})(v(0)\cdot
\hat{r}})}{c}], \end{eqnarray}

where ${\bf \hat{r}}$ is the unit vector towards the galaxy in question, 
${\bf v(r)}$ is the peculiar velocity of a galaxy at position ${\bf r}$,  
and {\bf v(0)} is the non-comoving velocity of the observer. 
 
To first order approximation and in the Local Group
rest frame, the distance of a galaxy (expressed in \ks) can be 
written as: 

    \begin{equation} \label{dist} r=cz-[{\bf v}(r,l,b)
-{\bf v}(0)]{\bf\cdot\hat {r}}
\end{equation}

where $l$ and $b$ are the galactic longitude and latitude. 

Thus, with a peculiar velocity field model which describes point by
point the function $ {\bf v}(r,l,b)$, it is possible to correct the
linear Hubble law in order to take into account peculiar motions. 
We rely on two basic approaches to modeling the peculiar velocity field. 
They are described below. 
 
\subsection{The Modified Cluster Dipole Model}

\subsubsection{The Model}

We use the BP96 optical cluster dipole model as a basic predictive model 
of the peculiar velocity field. The BP96 reconstruction scheme of the
3D positions and peculiar velocities of galaxy clusters is
based on the observed distribution of an optical cluster sample (with $r
\leq 25000$~\ks) extracted from the Abell/ACO sample (Abell, 1958; 
Abell, Corwin \& Olowin, 1989). Taking into account observational biases
through an homogenization procedure and filling artificially the zone of
avoidance with a simulated population of galaxies, BP96 obtained a
statistically homogeneous all-sky sample of clusters. 

Under some general and reasonable assumptions, e.g.  that the peculiar
velocities are caused by gravitational instability, that linear
instability theory applies and that cluster density fluctuations are
related to the mass density fluctuations by a constant linear biasing
factor $\delta_c({\bf r})=b_c\delta ({\bf r})$, 
a simple relation between the peculiar velocity of a galaxy
and the surrounding mass density field can be derived (cf Peebles, 1980):  

\begin{equation}\label{velin} {\bf v(r)=\frac{\beta}{4\pi}\int\delta (r^{'})
\frac{r^{'}-r}{|r^{'}-r|^3}d^3r^{'}},
\label{Peebles}\end{equation}

where $\delta (r)=[\rho (r)-\rho_b]/\rho_b$ is the density 
fluctuation about the mean background density $\rho_b$ and
$\beta=\frac{f(\Omega_{0})}{b_c}\sim \frac{\Omega_{0}^{0.6}}{b_c}$.

Within the linear theory approximation, BP96 used an iterative
reconstruction algorithm to remove redshift space distortions,
recovering both the real space distribution of clusters and  
the field of peculiar velocities generated by this cluster distribution. 

The resulting density and velocity fields clearly depend on an assumed
value of $\beta$, which is not directly provided by the procedure itself.
Nevetheless, this kind of analysis yields a resulting cluster 3D dipole
pointing only $\sim 10^{\circ}$ away from the CMB apex; therefore,
comparing the reconstructed cluster dipole amplitude to the LG peculiar
velocity (as inferred by the CMB dipole) BP96 estimated $\beta \sim$0.21. 
                                                                              
The model gives some important information (such as bulk flow amplitudes)
on large scales, but we are interested in constructing an accurate picture
of the velocity field and in recovering galaxy distances in the region
which lies within about 5500~\ks. 

Because of the heavy smoothing procedure used, the cluster density is not
known with a good local resolution. In other words, local contributions to
the velocity field such as those originating from the gravity of the Virgo
cluster, which is not even included in the Abell/ACO

\end{multicols}
$\ \ \ \ \ \ $\\
\vspace{10.5truecm}
$\ \ \ $\\
{\small\parindent=3.5mm {Fig.}~1.---
The plots show the radial components (in the CMB
frame) of the smoothed velocity fields in the supergalactic plane $SGX,
SGY$ for the cluster dipole model by Branchini \& Plionis (1996) (left)
and for the cluster dipole model modified with the inclusion of a
Virgocentric infall (right). The field is smoothed with a 1500 \ks
Gaussian window and is normalized to $\beta=0.21$. The arrows and the
boldface arrows distinguish between incoming and outcoming objects. The
contours correspond to the same radial peculiar velocity;  contour spacing
is 100 \ks, with the heavy contour marking 0. Regions of different radial
peculiar velocities are also indicated by shading. The Local Group is at
the center, the Great Attractor is on the left, Perseus-Pisces is on the
right, and Coma is at the top.  Coordinates are expressed in \ks.
\label{frad1}
}
\vspace{5mm}
\begin{multicols}{2}

cluster sample
because of its large angular size, are not taken into account. 

We have address this problem by merging the cluster dipole model 
with a simple model of the Virgocentric infall. 
In our model we assume that the Virgo mass is spherically 
symmetric and is radially distributed  with a King density profile
\begin{equation} 
 \label{rking} \rho (r)=\rho_{b} 
\displaystyle \bigg[1+A\Big(1+x^{2}\Big)^{-1.5}\bigg]
\end{equation}
where $x=r/r_c$, $r$ is the distance of a galaxy from the Virgo 
center,  $r_c$ is a smoothing length, and $A$ is a normalization 
mass factor.  Using eq. \ref{velin} we obtain

\begin{equation} \label{vpec}
{\bf v(r)}=-A\Omega^{0.6} r x^{-3}\bigg[\ln{\big(x+\sqrt{1+x^2}
\big)}-\frac{1}{\sqrt{1+x^{-2}}}\bigg]
{\bf \hat{r}}.
\end{equation}

where ${\bf \hat{r}}$ is the unit vector pointing outward Virgo; the
parameter $r_c$ plays the role of a smoothing length necessary for the
linear theory to hold. 

We evaluate the influence of the Virgo cluster on the dynamics of the LS
by fitting the predictions of our modified cluster dipole model with the
observed radial velocities of a sample of nearby galaxies having
redshift-independent distances. 

In this fit we fix the coordinates of Virgo cluster at $l=284^{\circ}$,
$b=74^{\circ}$, $r=1350\;$~\ks (Virgo distance) as in Faber \& Burstein
(1988), whereas we treat $C=A\Omega^{0.6}$ and $r_c$ as free
parameters of the model.  Necessary ingredients in the fitting procedure
are the radial components of peculiar velocities as predicted by the model
($v_{i}^{mod}$), the corresponding observed recession velocities ($cz_i$)
and the galaxy distances $r_i$. The model parameters are determined by
means of a $\chi^{2}$ minimization procedure; we minimize the quantity

\begin{equation} 
    \chi^2=\sum_i \frac{[cz_{i}-cz_{i}^{mod}(r_{i},{\alpha_{k}})]^2}{
	\sigma_i^2} \label{chi2}
\end{equation}

where $cz^{mod}$ is given by eq. \ref{dist} and depends on a number of
model parameters ${\alpha_{k}}$. The quantity $\sigma^{2}$ is the
quadratic sum of the uncertainties in the redshift measures, errors in the
predicted redshifts induced by distance uncertainties plus a term which
describes noise in the velocity field, under the assumption that  these
errors are independent.

\end{multicols}
$\ \ \ \ \ \ $\\
\vspace{10.5truecm}
$\ \ \ $\\
{\small\parindent=3.5mm {Fig.}~2.---
The plots show the velocity field in the CMB
frame for the cluster dipole model (left) and the modified cluster dipole
model (right). The vectors shown are projections of the 3D velocity field
in the supergalactic plane $SGX, SGY$. The contours correspond to the same
velocity vector modulus; contour spacing is 100 \ks, with the heavy
contour marking 100 \ks. Regions of different peculiar velocities are also
indicated by shading. The Local Group is at the center, the Great
Attractor is on the left, Perseus-Pisces is on the right, and Coma is at
the top. Coordinates are expressed in \ks. \label{fvect1}
}
\vspace{5mm}
\begin{multicols}{2}

 The last term is a value of dispersion associated
to non-linear and non-spherically symmetric motions of galaxies which lie
around collapsed systems and for which our models fail to predict the true
radial velocity. For the first term we assume the value of 50 \ks, for the
last we adopt the value of 200 \ks (e.g., Strauss \& Willick 1995).  We
estimate the formal $1\sigma$ errors of the model parameters from the
covariance matrix calculated at the point which minimizes eq. \ref{chi2}
in parameter space. 

For this fit we choose the A82 spiral sample described in \S 2. These
galaxies are distributed quite uniformily in space around the Virgo
cluster and therefore they are suitable for studying the Virgo infall in
detail. The forward TF relation used for estimating the distances of these
galaxies has a $rms$ scatter of $\sigma=0.47$~ mag (Willick et al., 1997a). 
Specifically, we consider the 158 galaxies of the A82 sample which have
distances smaller than 3000 \ks. This upper limit reflects the vanishing
of Virgo gravitational influence at large distances.  Moreover, the A82
sample is strongly incomplete at large distances, since it exhibits an
abrupt reduction in the number of objects per unit redshift at $cz_{\odot}
= 3000$ \ks. This redshift limit affects the reliability of the
inhomogeneous Malmquist bias corrections for the 59 objects lying beyond
that limit. 
    
\subsubsection{Model-fitting results}

The resulting best-fitting parameters $C=A\Omega^{0.6}$ and $r_c$ (\ks)
are tabulated in Table 1 together with the value of $\chi
^{2}$ per degree of freedom, which measures the goodness of the fit, and
the $rms$ dispersion of model velocity residuals $\sigma (cz)$ (in \ks).

\vspace{6mm} 
\hspace{-1mm}
\begin{minipage}{7cm}
\renewcommand{\arraystretch}{1.2}
\renewcommand{\tabcolsep}{1.2mm}
\begin{center}  
\vspace{-3mm}
TABLE 1\\
\vspace{2mm}
{\sc The best-fitting parameters for the modified cluster dipole model.\\}
\footnotesize
\vspace{2mm}
%
%
\begin{tabular}{cccccccc}                                                    
\hline
\hline
\multicolumn{2}{c}{C}                &                      
\multicolumn{2}{c}{$r_c$}            &
\multicolumn{2}{c}{$\chi^2/dof$}     &                                   
\multicolumn{2}{c}{$\sigma (cz)$}    \\
\multicolumn{2}{c}{}          &
\multicolumn{2}{c}{(\ks)}     &
\multicolumn{2}{c}{}          &
\multicolumn{2}{c}{(\ks)}     \\                                         
\hline                                                                
\multicolumn{2}{c}{$3.4\pm 1.3$} &\multicolumn{2}{c}{$810\pm 110$}&\multicolumn{2}{c}{$0.97$}&\multicolumn{2}{c}{$404$}  \\   

\hline
\end{tabular}

\end{center}
\vspace{3mm}
\label{fitplio}
\end{minipage}

Fig. 1 shows the smoothed radial velocity fields (in the
supergalactic plane $SGX,SGY$) of the BP96 cluster dipole model and our
modified cluster dipole model. Fig. 2 shows the smoothed 
velocity fields in the supergalactic plane $SGX, SGY$ for the same 
models. 

Although the BP96 cluster dipole model suffers from uncertainties in the
relation between fluctuations in matter and light distributions (e.g the
{\em linear biasing} assumption), from sample selection effects, and from
the method used to fill artificially the Zone of Avoidance, the model fits
fairly well the A82 data in the LS region.  The model implies a scatter of
$\sigma (cz)=485$~\ks (with $\chi^{2}/dof=1.44$) which can be compared
with the value of $\sigma (cz)=554$~\ks (with $\chi^{2}/dof=2.14$) which
one would obtain by fitting an unperturbed Hubble flow (in the CMB frame)
to the same data. 

The fit becomes even better if we add the perturbing presence of Virgo
cluster (see the low values of $\sigma (cz)$ and $\chi^{2}/dof$ listed in
Table 1). From our best fit we infer that the LG has a Virgo
infall velocity equal to $v=440\pm180$. This value is larger than those
quoted recently by several authors for whom the Virgo infall is not a
major source of the velocity field in the LS, in disagreement with earlier
contentions (see Han \& Mould, 1990 for a summary table). In recent years,
a large Virgo infall amplitude was also suggested by Tonry et al. (1992),
who, applying the surface brightness fluctuation technique as DI for local
ellipticals, found the value of $v=340\pm80$~\ks. On the other hand, from
the analysis of the magnitudes of bright cluster galaxies in a way which
is free of assumptions about motions in the LS, Gudehus (1995) found no
significant evidence of a Virgo infall of the LG.

The too great an amplitude of LG infall velocity is not due to
differences in the choice of observed Virgo redshift. It is instead likely
due to the fact that the BP96 cluster dipole model generates on the LG a
velocity vector (with projections along the supergalactic axes
$v_{SGX}=-225\pm61$ \ks, $v_{SGY}=131\pm40$ \ks, $v_{SGZ}=-349\pm81$ \ks)
which has a lower component in the general direction of GA
($v_{LG}=285$~\ks) than that predicted by other studies ($v \sim 500$~\ks;
see, e.g., Faber \& Burstein, 1988). So, in attempting to reproduce the
local observed flow pattern, our fitting parameters have to compensate the
underestimated GA infall of LG. This interpretation is consistent with the
fact that the CMB anisotropy dipole has a component directed towards the
Virgo cluster of $v=418\pm25$~\ks (see, e. g., the review by Davis \&
Peebles, 1983) and is supported by the fact that, by adding the Virgo mass
contribution, we find a velocity component of the LG motion towards GA of
$v_{LG}=558\pm170$~\ks, in good agreement with the value of $v_{LG}=535$~\ks
obtained by Faber \& Burstein (1988) with an independent model.

\subsection{The Multi-Attractor model}

\subsubsection{The Model}

The GA model of the Seven Samurai, proposed by Lynden-Bell et al. (1988)
and revised by Faber \& Burstein (1988), is the current most popular
phenomenological model of the large scale peculiar motions in the nearby
universe.  As claimed by several authors in the last years (e.g., Faber \&
Burstein, 1988; Han \& Mould, 1990; Shimasaku \& Okamura, 1992) and
substantially confirmed by a look at the mass density fields reconstructed 
through the application of the POTENT method to Mark II and Mark III  
data (e.g., Dekel et al. (1993) and Sigad et al. (1998), respectively),  
most of the local velocity field can be interpreted as
being generated by few gravitational sources characterized by spherical
symmetry. It is then sensible to describe the velocity field in term of a
multi-attractor toy model dominated by spherically symmetric Virgocentric
and GA-infalls.  In this model we consider also the possible effect of
other gravitational sources, such as the Perseus-Pisces Supercluster and
the Shapley concentration.  

This kind of model is an oversimplified {\em version of facts}, because it
imposes a simple, spherical geometry on the adopted gravitational sources
of a complex velocity field, which actually arises from a continuous field
of asymmetric density fluctuations.  Nevertheless, this model,
which as yet has never been applied to Mark III data, is the most
simple and statistically significant tool to analytically correct the
Hubble law.  As discussed below, the multi-attractor model yields
predictions in satisfactory agreement with data and it is adequate for our
purposes. 

Instead of considering the gravitational instability paradigm, we use the
infall model discussed by Reg\"os \& Geller (1989). In this model, based
on the Friedmann solution, the motion of galaxies at a particular radius
from the cluster center is approximated as the motion of a spherical mass
shell, which follows the same gravitational equation of motion as the
expansion factor of the universe. The shell can be treated as a Friedmann
universe on its own; it is possible to define a formal Hubble constant and
a formal cosmological density parameter $\Omega$ for the shell. This
infall model yields an implicit exact dependence, and an explicit
approximate analytic expansion, for the peculiar velocity as a separable
function of the present density contrast $\langle\delta_0\rangle$,
averaged inside a radius r, and the present cosmological parameter
$\Omega_0$. In the range \(0.1\leq\Omega_{0}\leq1\) this espansion can be
written (Reg\"os \& Geller, 1989) as: 

\begin{eqnarray} \label{infall}
\lefteqn{ {\bf v}_{p}=-{\bf r}
\bigg[\frac{1}{3}\Omega_{0}^{0.6}\langle\delta_{0}\rangle 
-0.063\Omega_{0}^{0.68}\langle\delta_{0}\rangle^2 +{} 
}
\nonumber\\
\displaystyle &+&      0.027\Omega_{0}^{0.71}
\langle\delta_{0}\rangle^3-0.015\Omega_{0}^{0.72}\langle\delta_{0}\rangle^4 
+{} 
\nonumber\\ 
\displaystyle
&+&     0.01
\Omega_{0}^{0.75}
\langle\delta_{0}\rangle^{5}
\cdots\cdots \displaystyle \bigg]  \end{eqnarray}

where the distance {\bf r} is expressed in \ks.

This series expansion provides convergence to the exact solution for
$\langle\delta_0\rangle\leq 1-2$. However because the linear term
dominates in the peculiar velocity, the $\Omega_0$ dependence of the exact
solution is close to that of the linear approximation. 

We approximate the total peculiar velocity as the sum of the 
components generated by the individual attractors. Assuming  
a King profile (eq. \ref{rking}) for each attractor (so that infall
velocities converge at the center of attracting mass), we build 
up a parameter toy model which is fitted to the various galaxy subsamples 
which are part of the Mark III data set.

Eq. \ref{infall} is strictly applicable only in the case of a single
attractor, as peculiar velocities can be added under the hypothesis, valid
in linear theory (and even beyond, see Susperregi \& Buchert, 1997) that
they are proportional to peculiar acceleration.  However, eq. \ref{infall}
reduces to linear theory almost anywhere, except in the proximity of an
attractor, where the contributions from all the other attractors are weak.
Therefore, the non-linear formula is considered as a suitable
approximation for the velocity field generated by the attractors. As a
consistency check, we have verified that the use of pure linear theory
does not change significantly the obtained values of the free parameters,
with the exception of $\Omega_{0}$, which can not be disentangled from the
normalization mass factor A of eq. \ref{rking} if no non-linearity is
taken into account. 
 
Specifically, we assume that the perturbing masses are: 1) the Virgo
cluster (V), which gives rise to distortions in the Hubble flow of LG
surrounding regions (e.g., de Vaucouleurs \& Bollingher, 1979; Tonry \&
Davis, 1981);  2) the Great Attractor (GA), which seems the best candidate
to explain large scale galaxy motions inside a volume of radius 5000-10000
\ks (Burstein et al., 1986; Dressler et al., 1987); 3) Perseus-Pisces
(PP), a large and irregular supercluster, whose massive filamentary
structure seems to be concentrated at a redshift of about 5000 \ks in the
antipodal direction of GA (Willick 1991; Han \& Mould 1992; Courteau et
al., 1993); 4) the Shapley concentration (SH), a very rich concentration
of $\sim$25 galaxy clusters whose mean complex lies at 14000 \ks and is
located behind the GA region (Scaramella et al., 1989). SH stands out as
the richest optical supercluster of the entire sky within $z<0.1$ (Zucca
et al., 1993). 

In our final analysis we do not consider the possible gravitational
influence of the Coma Supercluster, because we have checked that its
inclusion does not improve the goodness of the fit (see, e.g., 
Shaya, Tully \& Pierce, 1992, who reached the same conclusion in  
their less sensitive analysis of a galaxy sample limited to $cz=3000$ 
\ks).   

In the fit we do not consider the few very nearby galaxies having
$cz<700$~\ks and we do not correct their tabulated distances. These
galaxies are believed to share with the LG a bulk motion (the so-called
Local Anomaly) perpendicular to the supergalactic plane (Faber \&
Burstein, 1988). The Local Anomaly is a perturbation term introduced in
the standard picture of the peculiar velocity field of the very nearby
universe in order to explain the discrepancy between the observed and the
predicted radial velocities of the LG with respect to the CMB. It is in
general interpreted as a due to the "negative gravity" of the "Local Void"
(Tully \& Fisher, 1987) which is located above the supergalactic plane. 

After we have specified for each galaxy the redshift $cz_i$ and the
distance $r_i$, the $\chi^2$ expression (see eq. \ref{chi2}) becomes a known
function of several parameters. The free parameters which describe mass
distribution and geometry are, for each attractor, $A_i$, $r_{ci}$.
Furthermore, in the case of GA and SH, we have decided to leave free the
Galactic coordinates $l, b$ of their centers and also the distance of GA
from us. Finally, we try to leave $\Omega_{0}$ as a further free
parameter.  We take, instead, from the literature the values for the
positions of PP ($l=120^{\circ},\: b=-30^{\circ},\: r=5000$ \ks) and V
($l=284^{\circ},\: b=74^{\circ},\: r=1350$ \ks), and the distance of SH
($r=14000$ \ks). 

Thus, we are left with 13 degrees of freedom. Solutions may be unstable or
physically not meaningful in the presence of so many parameters.  However
our data set is very large and quite uniform in spatial distribution so
much to prevent statistical artifacts. Moreover, the improved accuracy in
estimated distances ensures confidence in the possibility of constraining
different models through the minimization procedure.  Lastly, if we assume
that the significance of each parameter is related to the change in
likelihood of the best fit that results from its addition to the model,
our trials with models having less parameters confirm the validity of the
best fits achieved with many degrees of freedom. Nevertheless, the
$\chi^2$ minimization scheme may introduce biased results due to the
incompleteness of the dataset. For example, the redshift limit of Mark III
catalog, which does not map distant regions of the universe, could 
constrain in a wrong way the parameters of the Shapley concentration.
 
The $\chi^2$ significance is ill-defined because velocity errors are
coupled and because it depends considerably on the size of the distance
errors and velocity spread due to noise in the velocity field. 
Encouragingly, we have checked that different values of the velocity
spread (taken as constant in space) have little influence on the attractor
parameters. In any case, we do not use the $\chi^2$ statistics to assess
the validity of a model.  We are interested in the relative decrease of
the $\chi^2$ values, because the purpose of our analysis is to obtain a
redshift-distance relation having less scatter and systematic bias than
that of the simple unperturbed Hubble flow.

\subsubsection{Model-fitting results}

Using various subsamples we estimate the attractor parameters separately
in order to examine their reliability and stability.
 
In Tables 2 and 3 we present the data number $N$, the
values of $\chi^{2}/dof$ and $rms$ dispersions of model velocity residuals
$\sigma(cz)$ (in \ks) relative to Mark II and Mark III data for Hubble
flows in the CMB frame and in the LG frame, respectively. 

We report results for the whole Mark II and for a Mark II subsample (Mark
II$^{*}$) which does not include the spirals by de Vaucouleurs \& Peters
(1984) and the field ellipticals.  The omission of the former subsample is
motivated by the fact that it does not appear in the subsequent
compilation (Mark III). The omission of the latter is justified by the
fact that many ellipticals reside in systems of fairly high velocity
dispersion and hence, can not be retained as good tracers of the velocity
field. In these tables we give also the results relative to the various
Mark III subsamples (denoted as described in \S 2), to the Mark III
spirals (Mark III$^{*}$) and the entire Mark III.

In Table 4 we give the minimization results (together

\clearpage

\end{multicols}
$\ \ \ \ \ \ $\\
\vspace{9truecm}
$\ \ \ $\\
{\small\parindent=3.5mm {Fig.}~3.---
The plots show the radial components (in the
CMB frame) of the velocity fields in the supergalactic plane $SGX, SGY$
for the multi-attractor model fitted to the Mark II subsample (left) and
the whole Mark III (right). The arrows and the boldface arrows distinguish
between incoming and outcoming objects. The contours correspond to the
same radial peculiar velocity; contour spacing is 100 \ks, with the heavy
contour marking 0. Regions of different radial peculiar velocities are
also indicated by shading. The Local Group is at the center, the Great
Attractor is on the left, and Perseus-Pisces is on bottom right. 
Coordinates are expressed in \ks.  \label{frad2}
}
\vspace{5mm}
\begin{multicols}{2}

\end{multicols}
$\ \ \ \ \ \ $\\
\vspace{9truecm}
$\ \ \ $\\
{\small\parindent=3.5mm {Fig.}~4.---
The plots show the velocity field in the CMB
frame for the multi-attractor model fitted to the Mark II subsample (left)
and the whole Mark III (right). The vectors shown are projections of the
3D velocity field in the supergalactic plane $SGX, SGY$. The contours
correspond to the same velocity vector modulus; contour spacing is 100
\ks, with the heavy contours marking 200 \ks and 300 \ks for the Mark II 
subsample and the Mark III, respectively. Regions of different peculiar
velocities are also indicated by shading. The Local Group is at the center
and the Great Attractor is on the left. Coordinates are expressed in \ks.
\label{fvect2}
}
\vspace{5mm}
\begin{multicols}{2}

\clearpage

\end{multicols}
\vspace{6mm} 
\hspace{-4mm}
\begin{minipage}{18cm}
\renewcommand{\arraystretch}{1.2}
\renewcommand{\tabcolsep}{1.2mm}
\begin{center}  
\vspace{-3mm}
TABLE 2\\
\vspace{2mm}
{\sc The Hubble flow model in the CMB frame.\\}
\footnotesize
\vspace{2mm}
\begin{tabular}{l@{\hspace{0.2cm}}c@{\hspace{0.2cm}}c@{\hspace{0.2cm}}c
@{\hspace{0.2cm}}c@{\hspace{0.2cm}}c@{\hspace{0.2cm}}c@{\hspace{0.2cm}}c
@{\hspace{0.2cm}}c@{\hspace{0.2cm}}c@{\hspace{0.2cm}}}
\hline
\hline
 &Mark II$^{*}$&Mark II&A82&MAT&HMCL&WCF&E/SO&Mark III$^{*}$&Mark III\\
\hline
${\cal N}$    &413 &662 &192 &277 &36 &65 &250 &570 &820  \\
$\chi^{2}/dof$&1.90 &1.70 &1.43 &1.16 &1.97 &1.30 &3.00 &1.22 & 1.97 \\
$\sigma (cz)$  &698 &835 &1020 &634 &696& 851 & 977 & 812 & 873 \\
\hline

\end{tabular}

\end{center}
\vspace{3mm}
\label{fitcmb}
\end{minipage}  
\begin{multicols}{2}

\end{multicols}
\vspace{6mm} 
\hspace{-4mm}
\begin{minipage}{18cm}
\renewcommand{\arraystretch}{1.2}
\renewcommand{\tabcolsep}{1.2mm}
\begin{center}  
\vspace{-3mm}
TABLE 3\\
\vspace{2mm}
{\sc The Hubble flow model in the Local Group frame.\\}
\footnotesize
\vspace{2mm}
\begin{tabular}{l@{\hspace{0.2cm}}c@{\hspace{0.2cm}}c@{\hspace{0.2cm}}c
@{\hspace{0.2cm}}c@{\hspace{0.2cm}}c@{\hspace{0.2cm}}c@{\hspace{0.2cm}}c
@{\hspace{0.2cm}}c@{\hspace{0.2cm}}c@{\hspace{0.2cm}}}
\hline
\hline
 &Mark II$^{*}$&Mark II&A82&MAT&HMCL&WCF&E/SO&Mark III$^{*}$&Mark III\\
\hline
${\cal N}$    &413 &662 &192 &277 &36 &65 &250 &570 &820  \\
$\chi^{2}/dof$& 2.10& 1.71& 1.22&1.49 & 1.96& 1.34&3.24 & 1.41   & 1.97  \\
$\sigma(cz)$  & 682& 790 & 943 &661  & 627& 1032&955 & 804   & 858  \\
\hline

\end{tabular}

\end{center}
\vspace{3mm}
\label{fitlg}
\end{minipage} 
\begin{multicols}{2}

\end{multicols}
\vspace{6mm} 
\hspace{-4mm}
\begin{minipage}{18cm}
\renewcommand{\arraystretch}{1.2}
\renewcommand{\tabcolsep}{1.2mm}
\begin{center}  
\vspace{-3mm}
TABLE 4\\
\vspace{2mm}
{\sc The best-fitting parameters for the multi-attractor model.\\}
\footnotesize
\vspace{2mm}
\begin{tabular}{l@{\hspace{0cm}}r@{$\pm$}l@{\hspace{0cm}}r
@{$\pm$}l@{\hspace{0cm}}r@{$\pm$}l@{\hspace{0cm}}r@{$\pm$}l@{\hspace{0.2cm}}r
@{$\pm$}l@{\hspace{0cm}}r@{$\pm$}l@{\hspace{0.2cm}}r@{$\pm$}l@{\hspace{0.2cm}}r
@{$\pm$}l@{\hspace{0cm}}r@{$\pm$}l@{\hspace{0cm}}r@{$\pm$}l@{\hspace{0cm}}}
\hline
\hline
 &\multicolumn{2}{c}{Mark II$^{*}$}&\multicolumn{2}{c}{Mark II}&
\multicolumn{2}{c}{A82}&\multicolumn{2}{c}{MAT}&\multicolumn{2}{c}{HMCL}&
\multicolumn{2}{c}{WCF}&\multicolumn{2}{c}{E/SO}&
\multicolumn{2}{c}{Mark III$^{*}$}&
\multicolumn{2}{c}{Mark III}\\
\hline
${\cal N}$&\multicolumn{2}{c}{413} &\multicolumn{2}{c}{662} &
\multicolumn{2}{c}{192} &
\multicolumn{2}{c}{277} &\multicolumn{2}{c}{36} &\multicolumn{2}{c}{65} &
\multicolumn{2}{c}{250} &\multicolumn{2}{c}{570} &\multicolumn{2}{c}{820}  \\
$\chi^{2}/dof$&\multicolumn{2}{c}{0.94}  &\multicolumn{2}{c}{0.99}&
\multicolumn{2}{c}{ 0.69} &\multicolumn{2}{c}{1.08} &\multicolumn{2}{c}{0.97}&
\multicolumn{2}{c}{ 0.78}&\multicolumn{2}{c}{2.60}&\multicolumn{2}{c}{0.98}&
\multicolumn{2}{c}{ 1.48} \\
$\sigma(cz)$  &\multicolumn{2}{c}{543}  &\multicolumn{2}{c}{ 686} &
\multicolumn{2}{c}{ 757}  &\multicolumn{2}{c}{618}  &\multicolumn{2}{c}{415} &
\multicolumn{2}{c}{ 739} &\multicolumn{2}{c}{903}&\multicolumn{2}{c}{713} & 
\multicolumn{2}{c}{787} \\
$A_V$& 3.1&0.6 & 2.9 &0.8  & 4.5 &0.8  &\multicolumn{2}{c}{ 0}   &
\multicolumn{2}{c}{0} &\multicolumn{2}{c}{0}  & 2.0 &1.2  & 4.0 &0.95  &
 3.1 &1.6   \\
$A_{GA}$& 2.9 &0.5   & 3.0 &0.7   & 3.7 &0.9  & 
\multicolumn{2}{c}{0}  & 3.2 & 1.0  & 3.9 &1.4  & 3.5 &0.7  & 1.5 &0.5  & 
2.4 &1.0  \\  
$A_{PP}$& 1.0 &0.7   & \multicolumn{2}{c}{-}  & 3.2 &1.7  & 
\multicolumn{2}{c}{0}  &
  4.0&1.2  & 2.8 &1.0  &\multicolumn{2}{c}{0}  & 3.4 &0.7  & 2.6 &1.1   \\  
$A_{SH}$& 1.0 &0.9   &\multicolumn{2}{c}{-}  &\multicolumn{2}{c}{0}   &  5.0&
 2.0  &
 4.2 &1.2  & \multicolumn{2}{c}{0}  & 2.0 &0.9  & 4.6 &1.4  & 3.3 &1.4   \\ 
$r_{c,V}$& 530 &30   & 500 &30   & 500 &28  & \multicolumn{2}{c}{0}  &
 \multicolumn{2}{c}{0}  & \multicolumn{2}{c}{0}  & 490 &44  & 480 &73  & 498 &
30   \\ 
$r_{c,GA}$& 2080 &45   & 1740 &70   & 1950 & 82  & \multicolumn{2}{c}{0}  &
 1830 & 70 &
 1930 &110  & 2500 &68  & 1820 & 112  & 1806 & 70   \\    
$r_{c,PP}$& 1480 &150   & \multicolumn{2}{c}{-}  & 1440 &110  &
\multicolumn{2}{c}{0}  &
 900 &50  & 2200 &110  &\multicolumn{2}{c}{0}   & 1410 &60  & 1415 & 61   \\ 
$r_{c,SH}$& 2450 & 170   & \multicolumn{2}{c}{-}     & \multicolumn{2}{c}{0}  &
  2500&130  &  3000 & 50  & \multicolumn{2}{c}{0}   &  2725& 80   & 2930 &
 50 &  3004 & 35    \\ 
$l_{GA}$& 305$^{\circ}$ & 4$^{\circ}$& 309$^{\circ}$ &4$^{\circ}$  & 
 294$^{\circ}$&
6.5$^{\circ}$  & \multicolumn{2}{c}{-}   & 314$^{\circ}$ &3$^{\circ}$  &
 303$^{\circ}$ &
3$^{\circ}$  & 309$^{\circ}$ &3$^{\circ}$ & 300$^{\circ}$ &18$^{\circ}$  &
 309$^{\circ}$ &5$^{\circ}$  \\     
$b_{GA}$& 15$^{\circ}$ &3$^{\circ}$  & 18$^{\circ}$ &3$^{\circ}$  &
 17$^{\circ}$ &
4$^{\circ}$ & \multicolumn{2}{c}{-}   &0$^{\circ}$&6$^{\circ}$ & 5$^{\circ}$&
3$^{\circ}$ & 10$^{\circ}$ &3$^{\circ}$  & 25$^{\circ}$ &19$^{\circ}$  &
 18$^{\circ}$ &3$^{\circ}$   \\     
$d_{GA}$& 4170 & 40  & 4170 &50   & 4150 &30  &  \multicolumn{2}{c}{-}  &
 4200 &90  &
 4200 & 40 &  4400&110  & 4200 &105  & 4200 &30   \\     
$l_{SH}$& 312$^{\circ}$ & 9$^{\circ}$  &   \multicolumn{2}{c}{-}  &
  \multicolumn{2}{c}{-}  & 319$^{\circ}$ &3$^{\circ}$ & 319$^{\circ}$ &
6$^{\circ}$  &
  \multicolumn{2}{c}{-}  &  308$^{\circ}$&4$^{\circ}$  & 308$^{\circ}$ &
 10$^{\circ}$ &
 308$^{\circ}$ & 5$^{\circ}$ \\     
$b_{SH}$& 27$^{\circ}$ &5$^{\circ}$  & \multicolumn{2}{c}{-}   &
 \multicolumn{2}{c}{-}&
  8$^{\circ}$&5$^{\circ}$  & 3$^{\circ}$ &4$^{\circ}$  &
 \multicolumn{2}{c}{-}   &
 2$^{\circ}$ &5$^{\circ}$  & 2$^{\circ}$ & 13$^{\circ}$ &  4$^{\circ}$&
 5$^{\circ}$ \\
 $\Omega_{0}$& 0.8 &0.2   &  1.0& 0.2  & 0.6 & 0.3 & 0.5 &0.3  & 0.5 &0.2  &  
0.8&0.3  & 0.2 &0.2  & 0.6 &0.2  & 0.4 &0.3   \\    

\hline
\end{tabular}

\end{center}
\vspace{3mm}
\label{fitma}
\end{minipage} 
\begin{multicols}{2}

with
$1\sigma$ errors) relative to Mark II and Mark III for our multi-attractor
model; the values of the GA distance $d_{GA}$, attractor radii $r_c$, and
model velocity residuals $\sigma(cz)$ are given in \ks. We have checked
that the inclusion of PP and SH did not improve significantly the fit to
the whole Mark II, so that only two attractors (Virgo and GA) are adopted
in this case.

For some parameters, in Fig. 5 we show the ellipses of
confidence (at the 68\% and 90\% significance levels) obtained from the 
covariance matrix, according to the model fitted to the whole Mark III. 
Fig. 3 shows the radial velocity fields (in the supergalactic
plane $SGX, SGY$) of the multi-attractor models relative to the Mark II
subsample and the whole Mark III. Fig. 4 show the
respective velocity fields. 

From an inspection of Tables 2, 3, and 4 we
can draw the following comments.  Velocity field models improve the fit
with respect to the unperturbed Hubble flow in a decisive way, and a
multi-attractor model is the best way to trace out the underlying observed
velocity field. Only for the MAT sample, a bulk motion generated by SH
seems to be best solution.  On the other hand, the fact that the uniform
expansion model in the LG frame generally fits the data much better then
that in the CMB frame strengthens the evidence for a large streaming
motion occuring along the PP-GA baseline and flowing towards SH. 

We stress that for most Mark III samples the best fit is obtained if SH is
included in the velocity field models. Only in A82 and WCF subsamples the
long range action of this supercluster is undetected, probably because the
former sample maps only the LS dynamics and the latter mostly covers the
region surrounding PP, which introduces incompleteness biases in the
$\chi^2$ statistics.

\includegraphics{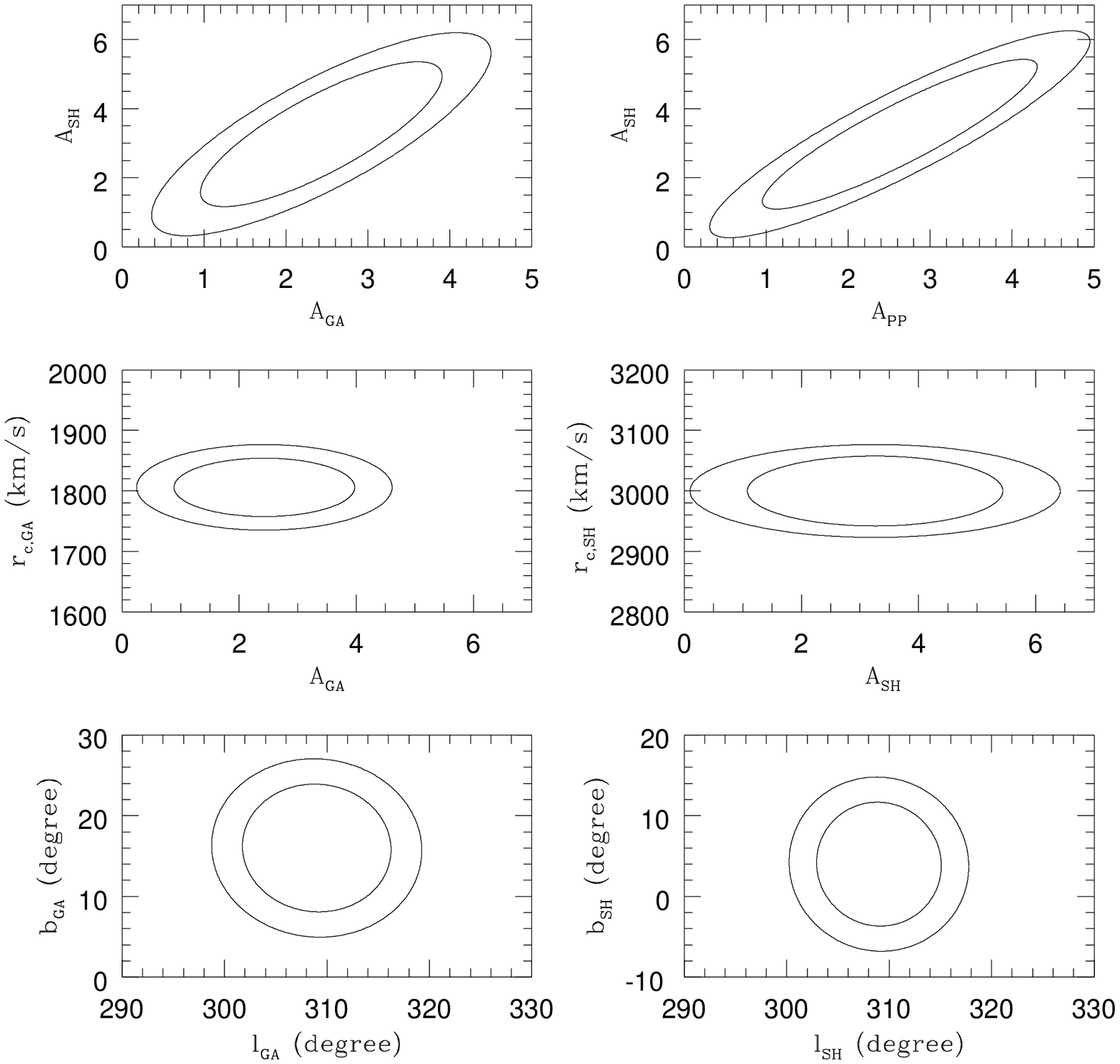}
$\ \ \ \ \ \ $\\
\vspace{8truecm}
$\ \ \ $\\
{\small\parindent=3.5mm {Fig.}~5.---
For some parameters, we show the regions of
confidence (at the 68\% and 90\% significance levels) according to the 
model fitted to the whole Mark III. 
\label{delta}
}
\vspace{5mm}

In every subsample the application of velocity corrections reduces
considerably both scatter and $\chi^{2}$. 

The reduction in scatter is
more evident in the A82, HMCL, WCF subsamples which have more accurate
distance estimates (see the relative $\sigma$-values in Tables
2, 3, and 4). Figs. 6,
7, and 8 show the recession velocity (in the CMB
frame) versus the distance before and after correction for peculiar
motions as predicted by models fitted to the three aforementioned
subsamples. These figures clearly illustrate how closer to the Hubble line
our corrected data lie than original data do. 

An independent way to judge the validity of a model is to look at the
velocity deviations from model predictions as a function of the distance.
Fig. 9 shows that, for the attractor models of the Mark II
subsample and whole Mark III, the residuals ({\it observed} minus {\it
calculated} velocity) exhibit no overall systematic variation with
distance, as is expected for a good model.

The position of the GA obtained from various Mark III subsamples is almost
similar and stable. The whole Mark II and Mark III yield $l=309^{\circ}$
$b=18^{\circ}$ for the GA as in Faber \& Burstein (1988). Our best fit of
the whole Mark III locates the mass center of SH at $l=308^{\circ}$
$b=2^{\circ}$, which somewhat differs from the position of the optical
center ($l=319^{\circ}$, $b=27^{\circ}$) as reported by Scaramella et al.
(1989). Interestingly, all Mark III subsamples tend to locate SH at low
$b$-values with respect to the optical position; this suggests that there
is more matter behind the Galactic plane than previously thought.  
Ongoing redshift surveys in the SH region (e.g., Proust, Quintana 
\& Slezak, 1998) will be able to investigate on the possible SH extension 
towards the zone of avoidance.  

The relative values of the radii $r_c$ give some insight into the
distribution of matter inside the attractors. The large $r_c$-values of GA
and SH indicate that they are not peaks of highly clumped matter, but
rather large overdensity regions. In particular, the GA mass is
substantially distributed across a distance of $\sim 2000$ \ks (Faber \&
Burstein (1988) quoted the value of 1500 \ks for a different model); the
PP core radius is $\sim 1400$ \ks; the SH core radius is $\sim 3000$ \ks,
whereas the optical radius, defined as the radius of a sphere centered in
SH and containing $\sim$25 (or $\sim$15) rich clusters, is $\sim$ 5000 
\ks (or $\sim$1500 \ks) (Vettolani et al., 1990; Bardelli et al., 1994). 

The normalizing mass parameters $A_i$, which are left as free parameters, 
bear no immediate interpretation. However, these values are used to
evaluate interesting physical quantities, such as the masses of the
attractors or the infall velocities that they generate. 

To estimate the excess mass in the attractor regions, we evaluate the
volume integral of the mass distribution we adopted (the King profile of
eq. (4)). This is a crude approximation because of the possible strong
departure of the mass distribution from spherical symmetry;  we suppose
that anisotropies in the density field cancel out when averages over a
large volume of space. For a sphere centered in GA and having LG on its
border, the whole Mark II yields a mass of $(5.1\pm2.5)\cdot10^{16}\cdot
h^{-1} M_{\odot}$, whereas Lyndell-Bell et al. (1988), using only the
elliptical sample, reported the value of $5.4\cdot 10^{16} M_{\odot}$, and
Shaya et al. (1992) reported $1.5\cdot\Omega_0^{0.4}\cdot 10^{16}
M_{\odot}$. The whole Mark III gives a smaller mass of 
$(1.8\pm1.6)\cdot10^{16}\;h^{-1}\;M_{\odot}$.

If we consider the attractors as spheres of radius $r_c$, we obtain the
excess masses reported in Table 5 for our multi-attractor models
fitted to some data subsets (Mark II subset, Mark III spirals, whole Mark
III). Table 5 lists also the density contrast averaged over a
sphere centered on the attractor and having the LG on its border.

According to Mark III and Mark II, Virgo is confirmed to be a poor
cluster; it has an excess mass of $1.3\cdot 10^{14} h^{-1} M_{\odot}$ (see
Table 5) and a total mass of $2.0\cdot 10^{14} h^{-1} M_{\odot}$
according to Mark III data ($2.2\cdot 10^{14} h^{-1} M_{\odot}$ and
$3.2\cdot 10^{14} h^{-1} M_{\odot}$ according to Mark III spiral data,
respectively). These values are substantial agreement with the estimates
of masses based on optical and X-ray data. From optical data Girardi et
al. (1998) estimated a virial mass of $M=(2.7^{+0.5}_{-0.4})\;10^{14}
h^{-1} M_{\odot}$ within a virialization radius of $1.69 h^{-1}$ Mpc. From
ROSAT PSPC X-ray observations Nulsen \& B\"ohringer (1995) estimated a
cluster mass per unit length of $1.24\cdot10^{11}\;M_{\odot}\;kpc^{-1}$,
which is 

\end{multicols}
\includegraphics{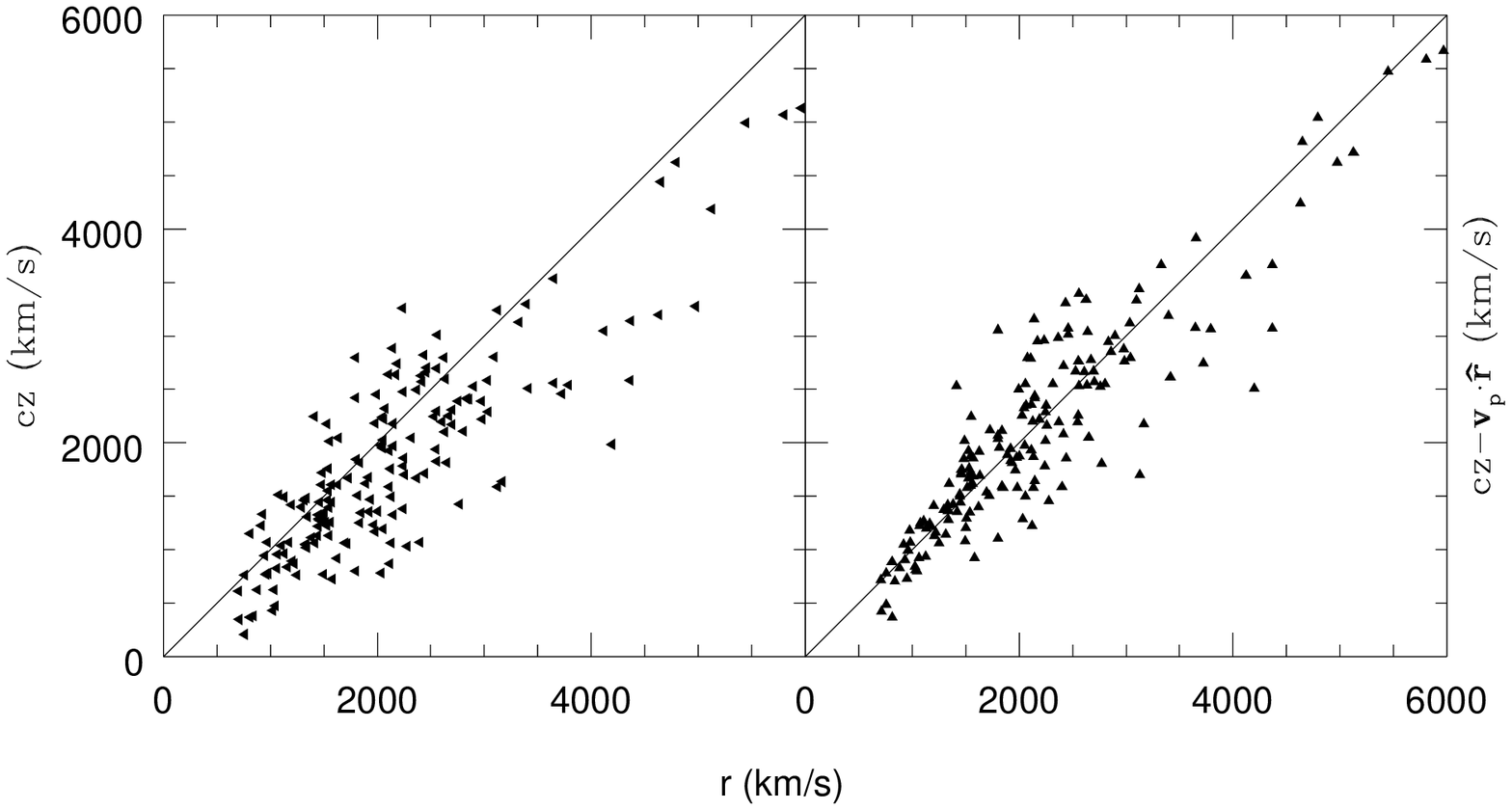}
$\ \ \ \ \ \ $\\
\vspace{10truecm}
$\ \ \ $\\
{\small\parindent=3.5mm {Fig.}~6.---
These plots show the recession velocity $cz$ (in
the CMB frame) as a function of the distance r (in km/s) before (left) and
after (right) corrections for peculiar motions, as predicted by the
multi-attractor models fitted to the A82 galaxy sample. The scatter
decreases from $\sigma=1020$ (left) \ks to $\sigma=757$ \ks (right).  
\label{conf1}
}
\vspace{5mm}
\begin{multicols}{2}

\end{multicols}
\includegraphics{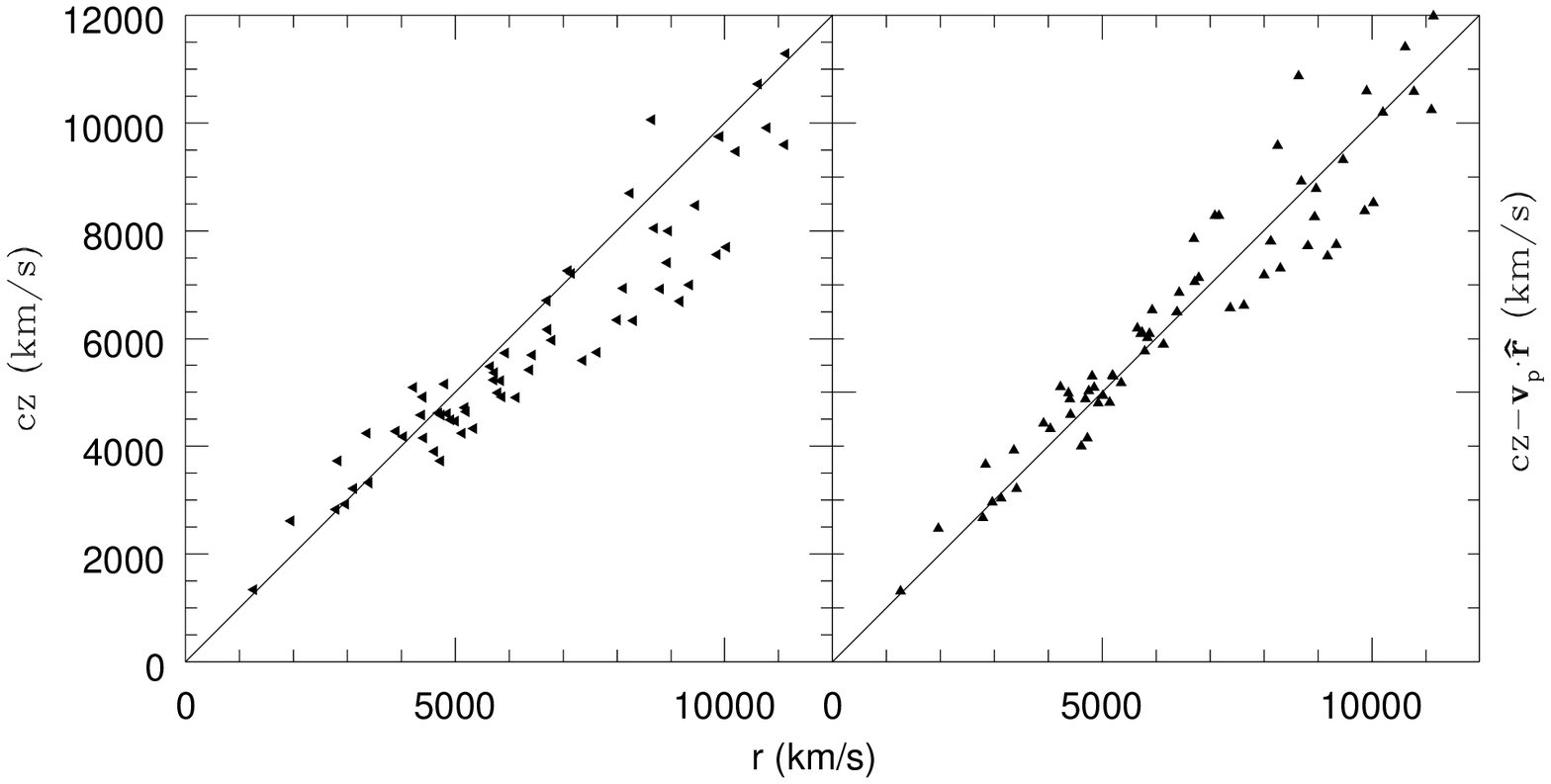}
$\ \ \ \ \ \ $\\
\vspace{9.5truecm}
$\ \ \ $\\
{\small\parindent=3.5mm {Fig.}~7.---
The same as in Fig. 6 for the
multi-attractor model fitted to the WCF sample. The scatter decreases from
$\sigma=851$ \ks (left) to $\sigma=739$ \ks (right). \label{conf2}
}
\vspace{5mm}
\begin{multicols}{2}

\clearpage

\end{multicols}
\includegraphics{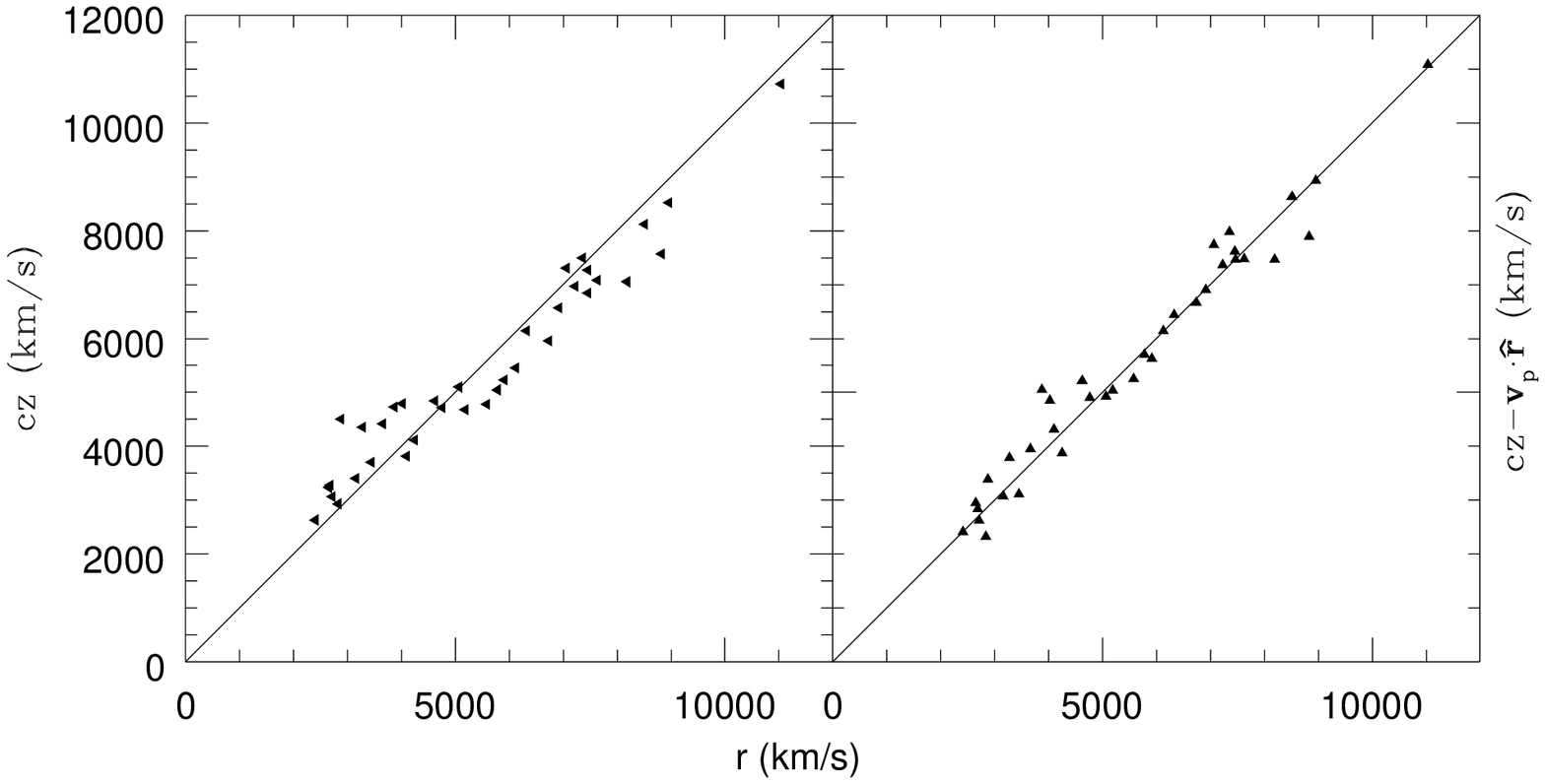}
$\ \ \ \ \ \ $\\
\vspace{9.5truecm}
$\ \ \ $\\
{\small\parindent=3.5mm {Fig.}~8.---
The same as in Fig. 6 for the
multi-attractor model fitted to the HMCL sample. The scatter decreases
from $\sigma=696$ \ks (left) to $\sigma=415$ \ks (right). \label{conf3}
}
\vspace{5mm}
\begin{multicols}{2}

consistent with our estimates, if we use an isothermal sphere
extrapolation ($M(r)\propto r$).
 
Compared to Mark II, Mark III data stress the role of PP and SH (see also
the $A_i$-values listed in Table 4), whilst they give a less prominent GA
with a mass which is roughly comparable to the PP mass and is $\sim$ 5
times smaller than the SH mass. This is in good agreement with Scaramella
et al.'s (1989) finding that in optical maps the SH region contains 5
times as many rich clusters as the GA region does. The SH excess mass
derived from Mark III ($3-5.7\cdot 10^{16} h^{-1} M_{\odot}$, see Table
5) (as well as the corresponding SH total mass of $4.7-8.1\cdot
h^{-1}\cdot 10^{16} M_{\odot}$) is roughly consistent with values
intermediate between upper and lower limits based on optical and X-ray
data.  From the velocity dispersion of the SH clusters Raychaudhaury et
al.  (1991) and Quintana et al. (1995) obtained upper limits to the total
mass of $1.3\cdot10^{17}h^{-1}\;M_{\odot}$ and
$7\cdot10^{16}h^{-1}\;M_{\odot}$, respectively.

 Assigning to the 40 SH
clusters the virial mass of $3\cdot 10^{14} h^{-1}\;M_{\odot}$, which is
the average of the virial masses deduced for 10 SH clusters, Quintana et
al. (1995) estimated a lower limit for the SH mass of $1.2\cdot10^{16}
h^{-1}\;M_{\odot}$.  From the X-ray luminosities of the 12 brightest
clusters, scaling to the Coma cluster (with $L_X\propto M^{0.4}$), 
Raychaudhury et al. (1991) obtained a lower limit of
$9\cdot10^{15}\;h^{-1} M_{\odot}$.  From the analysis of the ROSAT PSPC
and {\it Einstein Observatory} IPC X-ray observations Ettori, Fabian \&
White (1997) derived a lower limit for the dynamical SH mass of $6\cdot
10^{15} h^{-1}\;M_{\odot}$; their virial mass estimate is $\sim$5 times
larger than the last value.  From the SH influence on local flows Shaya et
al. (1992) reported a large dynamical mass of
$2.6\cdot\Omega_{0}^{0.4}\;10^{17}\;M_{\odot}$. The same authors obtained
a PP

\end{multicols}
\vspace{6mm} 
\hspace{-4mm}
\begin{minipage}{18cm}
\renewcommand{\arraystretch}{1.2}
\renewcommand{\tabcolsep}{1.2mm}
\begin{center}  
\vspace{-3mm}
TABLE 5\\
\vspace{2mm}
{\sc The excess masses and density contrasts for the attractors.\\}
\footnotesize
\vspace{2mm}
\begin{tabular}{lcccccc}                                                  
\hline \hline
\multicolumn{1}{c}{}&      
\multicolumn{2}{c}{Mark III}&
\multicolumn{2}{c}{Mark III(spirals)} &
\multicolumn{2}{c}{Mark II(subset)} 
\\
 Attractors       &                                                  
  $ M(<r_c)$              &                                          
  $\langle\delta_{0}\rangle_{LG}$  &                                 
  $ M(<r_c)$                 &                                      
  $\langle\delta_{0}\rangle_{LG}$ &             
  $ M(<r_c)$                  &                                      
  $\langle\delta_{0}\rangle_{LG}$                                   
\\                                                                            
  &                                                                  
 ($M_{\odot}h^{-1}$)& 
     &                                
 ($M_{\odot}h^{-1}$)& 
    &                                 
 ($M_{\odot}h^{-1}$)& 
  \\                                                                     
\hline                                                    
Virgo  &$(1.3 \pm1.3 )\cdot10^{14}$&$0.36 \pm 0.22$ &              
$(2.2\pm1.5) \cdot10^{14}$ &$0.44\pm0.23$ &                              
$(2.8\pm1.5 )\cdot 10^{14}$ &$ 0.35\pm0.30$ \\                     
GA&$(4.7 \pm4.4) \cdot10^{15}$&$0.38  \pm 0.18$&                
$(4.4\pm3.0)\cdot10^{15}$ &$0.24\pm0.11$ &                             
$(1.6\pm 0.6 ) \cdot 10^{16}$&$ 0.44\pm0.10$ \\                       
PP  &$(2.4 \pm2.2) \cdot10^{15}$&$ 0.19 \pm 0.09$ &                          
$(4.7\pm2.7)\cdot10^{15}$ &$0.23\pm0.06$ &                                    
$(1.5\pm2.1)\cdot10^{15}$ &$0.07\pm0.07$\\                                   
Shapley&$(3.0\pm2.7 )  \cdot 10^{16}$&$ 0.12 \pm 0.05$&        
$(5.7\pm3.7)\cdot10^{16}$ &$0.16\pm0.06$ &                                   
$(8.0\pm9.8)\cdot10^{15} $&$0.02\pm0.02$  \\                      
\hline                                                         
\end{tabular}                                              
\end{center}
\vspace{3mm}
\label{mass}
\end{minipage} 
\begin{multicols}{2} 

\clearpage

\includegraphics{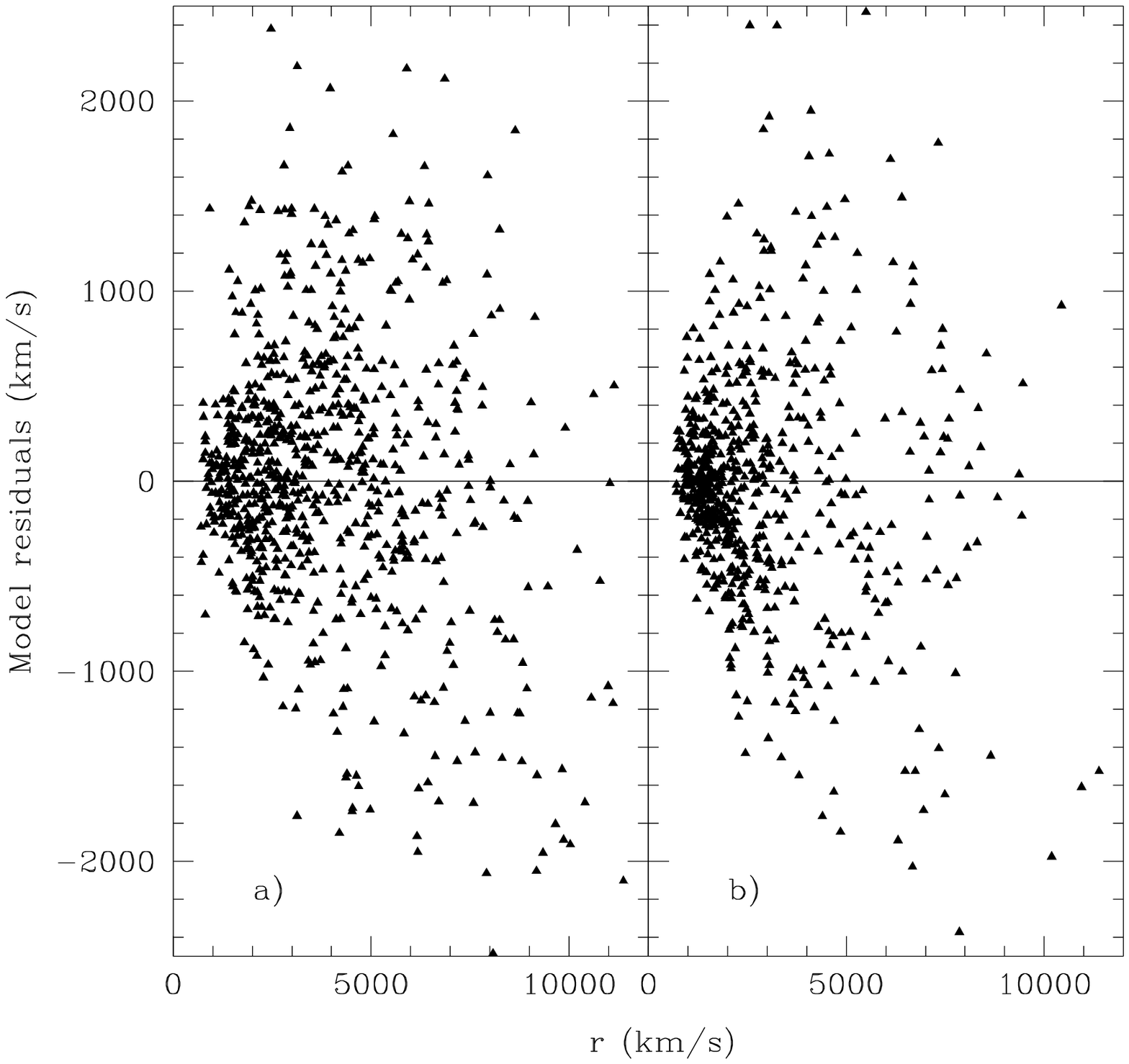}
$\ \ \ \ \ \ $\\
\vspace{8truecm}
$\ \ \ $\\
{\small\parindent=3.5mm {Fig.}~9.---
We show the residual ({\it observed}
minus {\it calculated}) velocities versus the distance $r$ (in \ks), for
the velocity models fitted to the whole Mark III ($a$) and the
Mark II subsample ($b$).  \label{residui}
}
\vspace{5mm}

dynamical mass of $5\cdot\Omega_{0}^{0.4}\;10^{16}\;M_{\odot}$, which
better agrees with our estimates than their SH mass does.  The
best-fitting values of $\Omega_{0}$ tabulated in Table 4 are
used for the estimates given above. 

We have also computed the infall velocity generated by Virgo alone
at the LG position. We find $v=170\pm64$ \ks, $v=145\pm57$ \ks, 
and $v=96\pm45$ \ks for the A82 sample, the Mark III spiral sample, 
and the whole Mark III, respectively. In particular the last low value 
indicates that Virgo infall is not a major source of the velocity 
field in the LS, in agreement with several recent results and 
in disagreement with the results of the modified cluster dipole 
model (see end of \S 3.1.2). 

Remarkably, the influence of the GA at the LG position is lower than
expected from Mark II and from previous results.

\vspace{6mm} 
\hspace{-1mm}
\begin{minipage}{7cm}
\renewcommand{\arraystretch}{1.2}
\renewcommand{\tabcolsep}{1.2mm}
\begin{center}  
\vspace{-3mm}
TABLE 6\\
\vspace{2mm}
{\sc The motion of the Local Group in the CMB frame.\\}
\footnotesize
\vspace{2mm}
                                                                              
\begin{tabular}{cccccc}    
\hline \hline
 Sample    &                                                         
 l         &                                                         
 b         &                                                         
 $\theta$  &                                                         
 $V$&                                                          
 $V_{\parallel}$        \\                                    
\hline                                                                    
                                                                              
Mark II(subset)&$ 305^{\circ}$ &$ 25^{\circ}$ &$ 25^{\circ} $& 705 & 567 \\    
Mark II&$ 307^{\circ}$ &$ 29^{\circ}$ &$ 26^{\circ} $& 666 & 563 \\           
A82&$ 290^{\circ}$ &$ 30^{\circ}$ &$ 11^{\circ} $& 479 & 615 \\               
MAT&$ 310^{\circ}$ &$ 7^{\circ} $ &$ 48^{\circ} $& 412 & 491 \\               
HMCL&$ 318^{\circ}$ &$ 0^{\circ} $ &$ 49^{\circ} $& 810 & 410 \\           
WCF&$ 310^{\circ}$ &$ 0^{\circ} $ &$ 43^{\circ} $& 225 & 456 \\               
E/SO&$ 309^{\circ}$ &$ 12^{\circ}$ &$ 34^{\circ} $& 556 & 516 \\              
Mark III(spirals)&$ 309^{\circ}$ &$ 7^{\circ} $ &$ 37^{\circ} $& 454 & 496 \\  
Mark III&$ 309^{\circ}$ &$ 12^{\circ}$ &$ 36^{\circ} $& 488 & 511 \\           
\hline                                                                   
                                                                  
\end{tabular}

\end{center}
\vspace{3mm}
\label{dip}
\end{minipage}  

The pull generated by GA
at the LG position turns out to be $v=314\pm200$ \ks according to Mark
III data (it is $v=620\pm260$ according to Mark II data). Therefore,
nearly one half of the LG motion is due to the SH action. At variance with
Mark II, Mark III data suggest that the SH, projected behind the GA,
yields a pull on the LG comparable to that exerted by GA. 

There is a way to test the relative prominence of the GA without resorting
to any specific attractor modeling. It suffices to assume that the infall
on the attractor be radial and that we know fairly well its angular
position in the sky. Let us take a sphere centered at half the distance
between LG and GA, with LG on its border. Now if the infall is radial and
GA is the only responsible for the observed flow patterns, then, for a
simple geometrical reason, the galaxies lying inside the sphere should
have positive peculiar velocities, whilst the galaxies lying outside
should have negative radial components (see Fig. 10); only at the
surface of the sphere the radial components would become null.

\includegraphics{fig10.eps}
$\ \ \ \ \ \ $\\
\vspace{8truecm}
$\ \ \ $\\
{\small\parindent=3.5mm {Fig.}~10.---
The plot illustrates the following concept: for a 
sphere centered at half the distance between the Local Group and 
the Great Attractor, under the hypothesis of a radial infall on the Great 
Attractor only, a galaxy which is at distance $d_c$ from the 
center of the sphere and lies inside (outside) the sphere should have a 
positive (negative) radial peculiar velocity $v_p$ (as viewed from 
an observer placed in the Local Group). \label{ga}
}
\vspace{5mm}

 In Fig. 11 we compare the results obtained with Mark II and Mark III
data. Mark II and A82 data (see Fig.  11d, f, respectively)
favour a massive GA dominating large scale motions, whilst evidence for
this is considerably weakened in other Mark III samples, which do not show
a clear radial infall pattern towards GA. This can be only partially due
to the perturbing presence of PP, because if we exclude the galaxies
located in the PP region from the plot (see  

\clearpage

\end{multicols}
\includegraphics{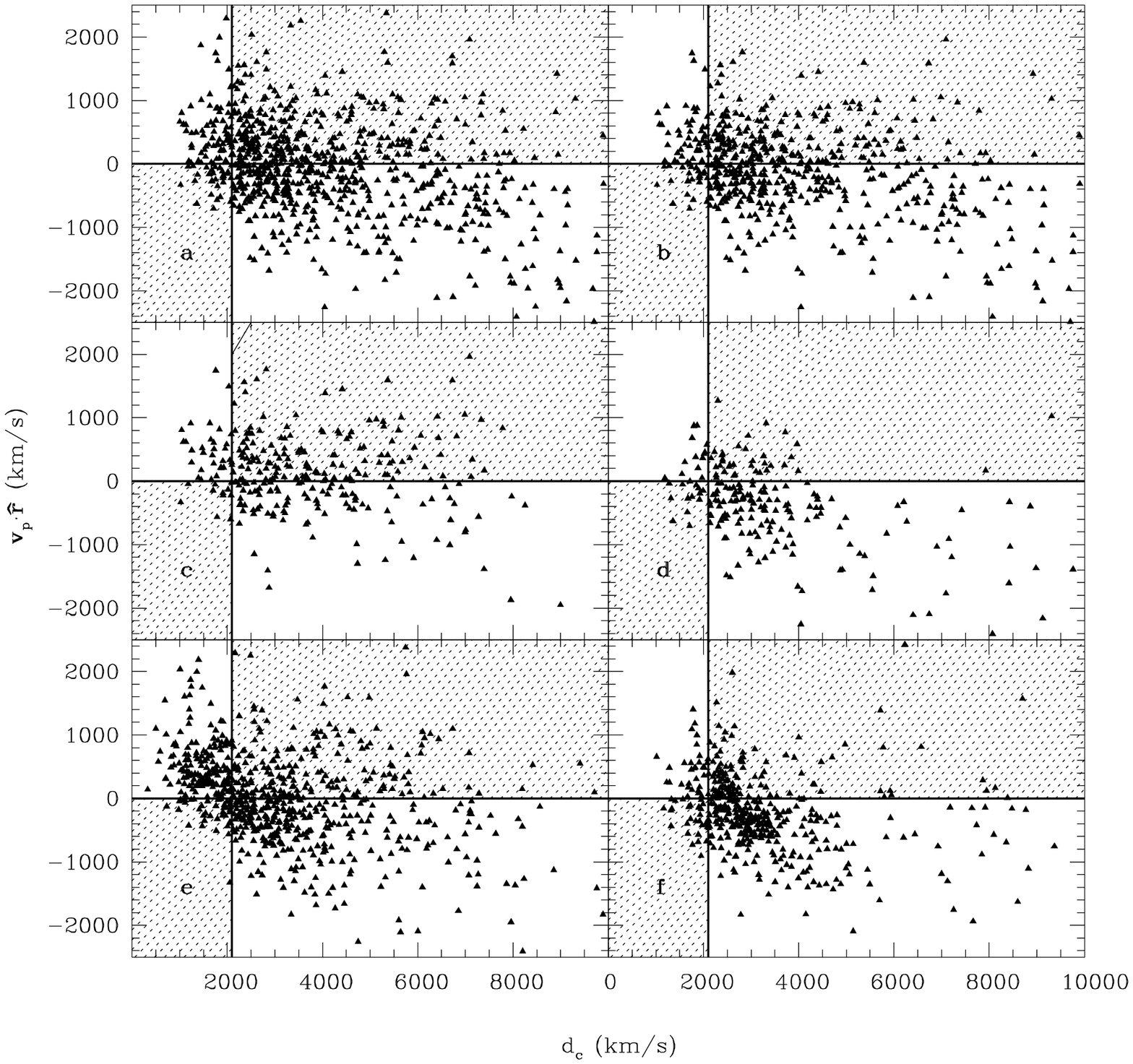}
$\ \ \ \ \ \ $\\
\vspace{17.3truecm}
$\ \ \ $\\
{\small\parindent=3.5mm {Fig.}~11.---
The observed radial peculiar velocities $v_p$ (in
\ks) as a function of the distance $d_c$ (in \ks) from the middle point of
a line joining the Great Attractor with the Local Group. If the Great
Attractor were the only source of radial infall, then for $d_c > 2100$ \ks
peculiar velocities should be negative, while for $d_c < 2100$ \ks they
should be positive (see Fig. 10).  We consider various samples of
galaxies: the whole Mark III data ($a$), the Mark III spirals ($b$)
(almost all Mark III ellipticals lie in the GA region), the MAT 82 spirals
($c$) , the A82 spirals ($d$).  the Mark III sample without the objects
located in the Perseus-Pisces region (i.e., those placed inside a sphere
centered on this supercluster and having a radius of 3000 \ks.) ($e$), the
whole Mark II data ($f$).  \label{cerchi}
}
\vspace{5mm}
\begin{multicols}{2}

Fig. 11e), there are
still many galaxies with positive peculiar velocities outside the sphere.

This reflects the lack of a back-side infall towards GA or, in other
words, the long range effect of SH which prevents galaxies located in the
back of GA to acquire negative peculiar velocity components.

If the GA
were the dominant system we would expect a GA back-infall to be comparable
in amplitude to the GA forward-infall. On the contrary, if SH exerts a
significant gravitational influence, we expect an asymmetry in the
amplitudes of the back- and forward-infalls, i.e. positive components
outside the sphere. In conclusion, local samples such as the A82 sample
and Mark II, which does not include many galaxies in the back of GA, can
not fully reveal the importance of SH in determining peculiar motions.

In Fig. 12 we show the mean overdensity profile inside a
spherical volume centered on each attractor, for our model of the whole
Mark III.  The PP and GA overdensities at $r\sim1200$ \ks 
are in reasonable agreement (being just a bit higher) with 
the density peaks at 1200 \ks Gaussian smoothing, ($\delta_{0}\sim$1.2
and $\sim$1.4, respectively) found by Sigad et al. (1998) who used the
POTENT procedure to reconstruct the smoothed mass density field from Mark
III data (for $\Omega_0$=1), taking into account mild non-linear effects.
These mass density peaks are not much different from the reconstructed 
real-space density peaks (at 1200 \ks Gaussian smoothing) of IRAS galaxies 
($\delta_{0}\sim$0.8; see Sigad et al., 1998) and optical galaxies 
($\delta_{0}\sim$1.8 and 0.8, respectively; see Hudson et al., 1995), 
which is in line with the widespread contention that mass and galaxies are 
related via an approximate linear biasing relation with a biasing factor 
$b_c$ of order unity.

In Table 6 we report the direction ($l, b$) and the amplitude $V$
of the CMB anisotropy dipole generated by the assumed distribution of
attracting masses, for multi-attractor models relative to various
subsamples.  We also give the angular separation $\theta$ between the
observed and reconstructed dipole together with the projection
$V_{\parallel}$ of the observed CMB dipole in the ($l, b$) direction.  In
all cases the reconstructed dipole shows a satisfactory agreement both in
direction and in amplitude with the observed dipole.

\includegraphics{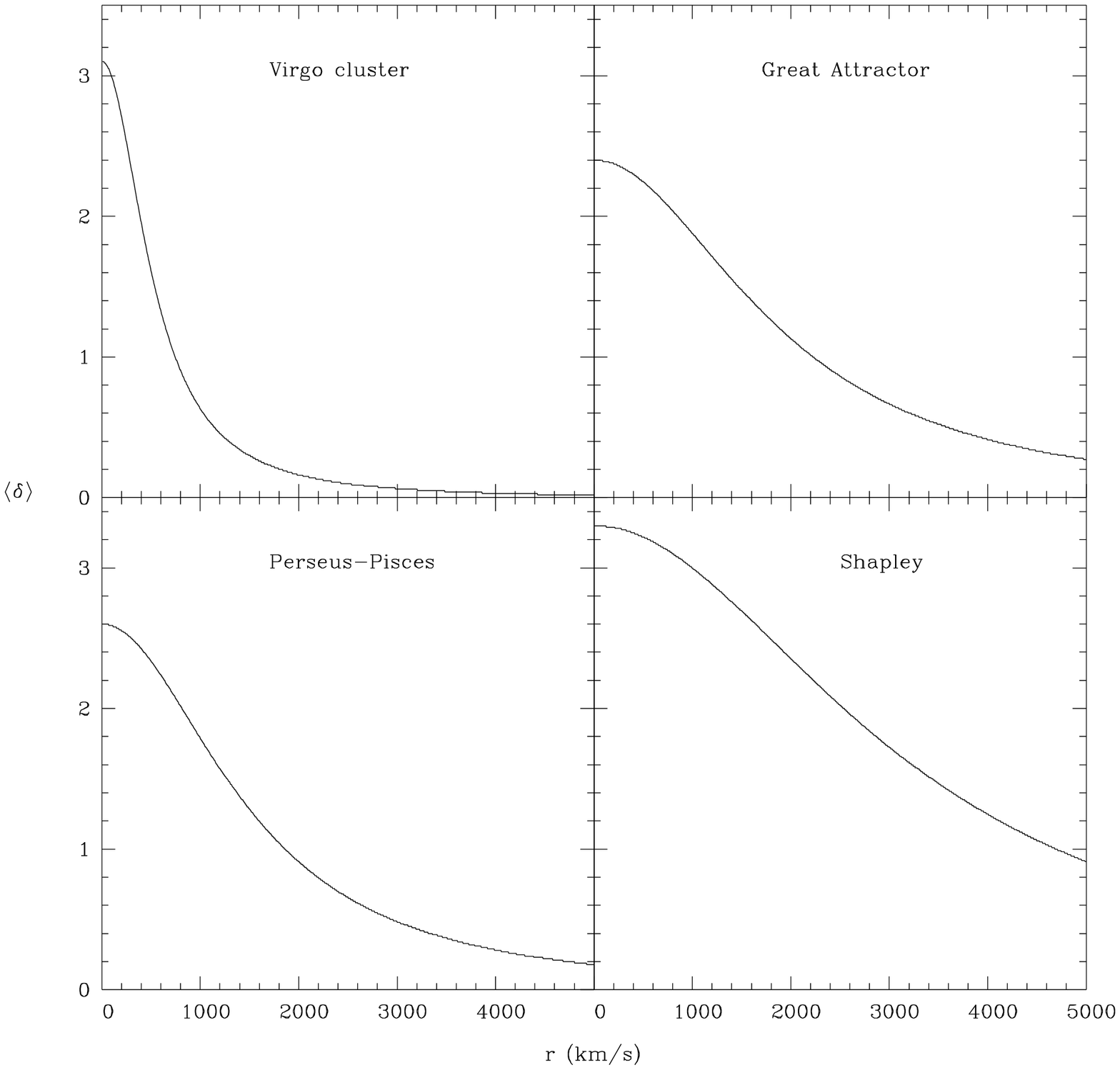}
$\ \ \ \ \ \ $\\
\vspace{8truecm}
$\ \ \ $\\
{\small\parindent=3.5mm {Fig.}~12.---
The plots show the density contrast,  
averaged over spheres centered on each attractor, as a function of the 
distance $r$ (in \ks) from the attractor, for the velocity model fitted to 
the whole Mark III data. \label{overdensity}
}
\vspace{5mm}

This can be taken as
an independent evidence of the reliability of the multi-attractor toy
model we used. 

The remaining differences between the observed and reconstructed dipole
can be justified in several ways. For example, inhomogeneities beyond
15000 \ks could contribuite significantly to the LG motion (e.g.,
Branchini, Plionis \& Sciama, 1996; Plionis \& Kolokotronis, 1998) or,
according to an alternative explanation, the motions of the very nearby
galaxies are also subjected to the Local Anomaly (Burstein, 1990). If we
preferred the latter explanation, we would reconstruct a Local Anomaly
velocity vector with an amplitude of 370 \ks pointing in the direction
$l=212^{\circ}$, $b=35^{\circ}$, which is not much different from that
reported by Faber \& Burstein (1988) ($v=360$ \ks, $l=199^{\circ},\;
b=0^{\circ}$) and is greater in amplitude than that obtained by Han \&
Mould (1990) ($v=240$ \ks, $l=205^{\circ},\; b=11^{\circ}$). 

In our minimizing scheme Mark II data favour $\Omega_{0}$-values close to
1, whereas Mark III subsamples tend to give lower $\Omega_{0}$-values (the
whole Mark III and the Mark III spirals give $\Omega_{0}=0.4\pm0.3$ and
$\Omega_{0}=0.6\pm0.2$, respectively).  Although the exact estimate of
$\Omega_{0}$ may be seriously affected by our specific modeling of the
density field in which the detailed features of density fluctuations are
neglected, the tendency towards lower $\Omega_{0}$-values is meaningful
and agrees qualitatively with the coming down of the estimates of the
parameter $\beta$ resulting from the comparison between the (Mark II and
Mark III) observed velocity field and that predicted by the IRAS galaxy
densities ($\beta\sim$0.4--0.9, with the lowest values favored by the most
recent works; see, e.g., Sigad et al., 1998). Notably, a comparison
between the IRAS galaxy density fields and the mass density fields
reconstructed from the application of the POTENT method to Mark II (Dekel
et al., 1993) and Mark III (Sigad et al., 1998) data shows a similar
tendency, with systematically greater values of $\beta$ ($\beta\sim$1.3
and $\sim$0.9, as reported in the respective papers).  Reasons for the
systematic differences in $\beta$-values obtained from comparing fields at
density and velocity levels have been discussed by Willick et al. (1997b)
and Willick \& Strauss (1998).  Direct comparison of densities is a local
procedure, whereas comparison of velocities is non-local, because the
velocity field is sensitive to the mass distribution in a large volume,
even outside the sampled volume (e.g., the Shapley concentration). This
non-local character is also inherent to our approach and may cause our
results to better agree with those resulting from velocity--velocity
comparisons.

Also very recent papers, which followed this kind of approach, favored
fairly low values of $\beta$, by using new samples of velocity field
tracers. Da Costa et al. (1998) found $\beta\sim$0.6 from an extensive
I-band TF sample of spiral galaxies, the SFI catalog (e.g., da Costa et
al., 1996).  Riess et al. (1997) obtained $\beta\sim$0.4 from a sample 
of Type Ia supernovae. 

In conclusion, our best estimate of $\Omega_{0}$ converges towards the
(inelegant) value of $\sim$0.5, which roughly corresponds to the average
of many estimates of $\Omega_{0}$ resulting from the analyses of large
scale structure and cosmic flows. Studies based on
non-linear dynamics within galaxies, groups, and clusters 
(on scales of $\sim 1-10\;h^{-1}$ Mpc) yield low values
of $\Omega_{0}\sim$0.2-0.3 (see the review by Dekel, Burstein \& White,
1997). 

\section{Inverting the redshift-distance relation: the triple-valued 
regions.}

We use three basic models derived in the previous section, i.e. the
modified cluster dipole model and the four-attractor models relative to
the Mark II subset and the Mark III spirals, for providing numerical and
explicit expressions of the recession velocity $cz$ (expressed in the LG
frame) as a non-linear function of distance $r$. In general, this
relation, coupled with the heliocentric redshift tabulated in Garcia et
al. (1993) and transformed by us in the LG frame according to the
relation 
$cz_{LG}=cz_{\odot}-79 \cos l \cos b + 296 \sin l \cos b - 36 \sin b$ 
(with $cz$ in \ks) (Yahil, Sandage \& Tamman, 1977), gives the
galaxy distance. The galaxy groups are given the median values of the
coordinates and redshifts of the group members, selected according to the
final catalog by Garcia (1993). As already said in \S 3.2.1, because of
the Local Anomaly problem, for the four-attractor model we do not invert
the redshift-distance relation for the few very nearby galaxies ($cz<700$
\ks), for which we simply take the tabulated distances (with $H_0=75\; km\;
s^{-1} Mpc^{-1}$). 
 
It is known that in the vicinity of prominent overdensities the
redshift-distance relation can become non-monotonic, such that there are
three different distances corresponding to a given redshift (see the
S-shaped curve of Fig. 14a in the following section). This makes
ambiguous the distance assignments for the galaxies which fall in these
triple-valued regions.

We treat as triple-valued zone objects also the field galaxies and the
groups which would fall in these zones, if their redshifts were 
modified by $\pm$100 \ks and by a value equal to the group velocity
dispersion, respectively. We calculate the velocity dispersions of groups
using the robust scale estimator defined in terms of the "median absolute
deviation" (Beers, Flynn \& Gebhardt, 1990). In our cases all
triple-valued zone field galaxies are spirals. 

In practice, if we take the modified cluster dipole model, this problem
concerns only 4 groups and 14 field galaxies in the Virgo region. If we
take the attractor model relative to the Mark II subsample, this concerns 10
groups and 28 field galaxies located in the Virgo or GA regions.  For the
attractor model fitting the Mark III spirals, the redshift-distance relation
remains monotonic in the vicinity of GA and there are 10 field galaxies
(and the Virgo cluster) located in the triple-values regions of Virgo or
PP. 
  
In order to solve this problem, Tully \& Shaya (1984) simply
relied on some DIs, optical appearance, and connection with some
Virgo clouds to guess the distance of the Virgo triple-valued zone
galaxies. Yahil et al. (1991) used a distance-averaging procedure, 
which however, tends to place all objects in the middle branch of the 
triple-valued zone; on the other hand, iteration schemes for  
translating from  redshift space to distance space  
tend to place objects outside the middle branch (Yahil et al., 1991; 
Hudson, 1993a). Sigad et al. (1998) used a statistical approach which takes 
into account prior information on the peculiar velocity field.

We use the following precepts to solve the problem of the triple-valued
zones. If a galaxy which falls in these zones is included in Mark III,
among the three possible distances we choose that which is closer to the
value given by Mark III. If a spiral galaxy is not included in Mark III,
but has known apparent magnitude $B_T$ and known maximum rotational
velocity $V_m$, we can evaluate the three values of the absolute magnitude
\begin{equation} M_B - 5 \log h= B_T - 5 \log r - 15.62 \end{equation}
which correspond to the three choices of distance r (expressed in \ks). 
Then we compare the values of $V_m$ and the three values of $M_B - 5 \log
h$ with the TF relation calibrated on a sample of galaxies lying at
similar redshift, but outside the triple-valued zone.  More precisely, for
the calibration of the TF relation expressed as the linear relation
\begin{equation} M_B - 5 \log h = a - b \log V_m \end{equation} (with
$V_m$ in \ks), we consider a sample consisting of all the galaxies which
lie in a distance shell centered on the attractor, with the shell having
the same width in distance as the distance range in which the
redshift-distance relation is not monotonic. We find the TF relation by
means of a regression analysis in which we calculate the 
ordinary least-squares bisector line, excluding 
iteratively the points which depart by 3$\sigma$ from the best regression
line. After having derived the TF relation, for a galaxy located in the
triple-valued zone we choose the distance which implies a value for the
absolute magnitude closer to the TF relation.

We wish to point out that in this procedure we do not want to find the
"true" TF relation to be used as an unbiased DI with low magnitude 
scatter; we simply use the TF as a criterium for choosing the distance of
a triple-valued zone object, which is found only in restricted volumes
close to prominent gravitational sources. In these circumstances
systematic biases act in the same way on all galaxies and therefore should
not affect our distance choices.

Table 7 reports the parameters of the TF relations (slope $b$,
zero-point $a$, $rms$ scatter $\sigma$ (in mag)) we used for various
regions and the three aforementioned peculiar velocity models. The slopes
are in substantial agreement with the typical values ($b\sim$5--6) 
reported in the literature for blue TF relations (e.g., Tully \&
Fisher, 1977; Bottinelli et al., 1983; Garcia et al., 1993).  As is
expected, the rms scatters $\sigma$ about the mean relations tend to be
larger than the typical value of $\sigma\sim 0.5-0.6$ mag for blue TF
relations (Tully \& Fisher, 1977), because of errors with which velocity
field models reproduce the distance, and, especially, because of the
inclusion in the sample of objects with inaccurate photometric and line
width data (e.g., galaxies with uncertain inclination corrections of the
21 cm line widths).

\end{multicols}
\includegraphics{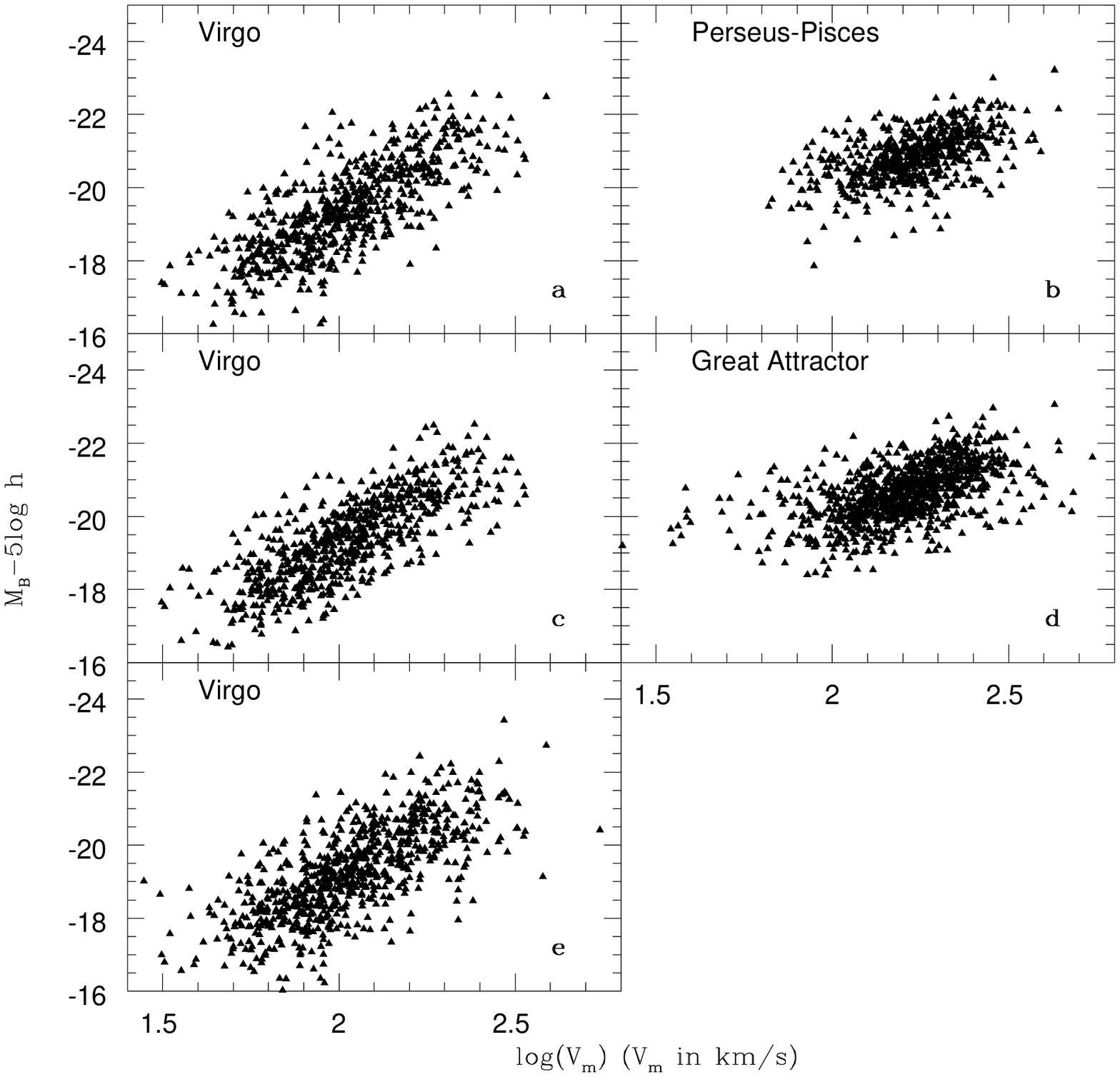}
$\ \ \ \ \ \ $\\
\vspace{13truecm}
$\ \ \ $\\
{\small\parindent=3.5mm {Fig.}~13.---
Plots of $M_B - 5 \log h$ (where $M_B$ is the blue
absolute magnitude) versus $\log V_m$ (where $V_m$ is the maximum rotation
velocity expressed in \ks) for the galaxies located in the Virgo redshift
region (e), (with galaxy distances predicted by the modified cluster
dipole model), for the galaxies located in the Virgo (c) and GA (d)
redshift regions (with galaxy distances predicted by the model fitted
to the Mark II subset), for the galaxies located in the Virgo (a)
and Perseus-Pisces (b) redshift regions (with galaxy distances
predicted by the model fitted to the Mark III spirals).  \label{TF}
}
\vspace{5mm}
\begin{multicols}{2}

 The substantial reliability of the velocity field
models considered is supported by the fact that they in general imply
reasonable TF relations.

Fig. 13 shows the $M_B - 5 \log h$ versus $\log V_m$ plots for galaxies
located in various redshift intervals corresponding to the the
triple-valued zones, with distances predicted by different velocity
models. 

If we had also field ellipticals (which we do not have) lying in the
triple-valued zones, we could have followed a similar procedure, using the
modified Faber-Jackson relation instead of the TF relation.  Remarkably, 
for the few triple-valued zone objects included in Mark III, our 

\end{multicols}
\vspace{6mm} 
\hspace{-1mm}
\begin{minipage}{17.7cm}
\renewcommand{\arraystretch}{1.2}
\renewcommand{\tabcolsep}{1.2mm}
\begin{center}  
\vspace{-3mm}
TABLE 7\\
\vspace{2mm}
{\sc The parameters of the Tully-Fisher relation.\\}
\footnotesize
\vspace{2mm}
\begin{tabular}{lrcccc}                                                    
\hline \hline
 Model    &                                                         
 Region         &
 {\cal {N}}  &
 a  &
 b         &                                                         
 $\sigma (mag)$    \\       
\hline
Multi-attractor (Mark III$^{*}$)&Virgo&641&-6.70 $\pm$0.35 &
6.27$\pm$0.16&0.83  \\
Multi-attractor (Mark II$^{*}$) &Virgo&665&-7.34 $\pm$0.32 &
5.93$\pm$0.25&0.79  \\
Modified cluster dipole model    &Virgo&743&-7.28 $\pm$0.40 &
5.87$\pm$0.20&0.94  \\ 
Multi-attractor (Mark III$^{*}$)&PP   &580&-11.66$\pm$0.46 &
4.11$\pm$0.21&0.63  \\   
Multi-attractor (Mark II$^{*}$) &GA   &957&-13.15$\pm$0.38 &
3.39$\pm$0.18&0.70  \\
\hline                                                                      
\end{tabular}

\end{center}
\vspace{3mm}
\label{ttf}
\end{minipage}  
\begin{multicols}{2}

 method for 
choosing the best distance yields the distance closer to 
the value tabulated in Mark III.

\section{Comparing the results relative to different velocity field
models}

In Fig. 14 we show the predicted behaviors of the recession
velocity and peculiar velocity against the distance from the attractor
center, along the lines of sight of Virgo, GA, and PP. We also show the
behavior of the peculiar velocity along the line of sight of SH. Fig.
14 and the previous Figs. 1, 2, 4, 5, well illustrate the main differences between the models.

There are outstanding features common to the velocity field models 
considered, such as the presence of the attractors Virgo, GA, PP and 
SH; but their prominence appreciably differs among the models 
(see the above-mentioned plots and \S 3.2.2). 

Notwithstanding the simple geometry adopted for the attractors, our
multi-attractor model fitted on Mark III data delineates a complex
velocity field, which well resembles the Mark III--POTENT velocity field
presented by Dekel (1994, 1997), except for the Coma supercluster region.

\end{multicols}
\includegraphics{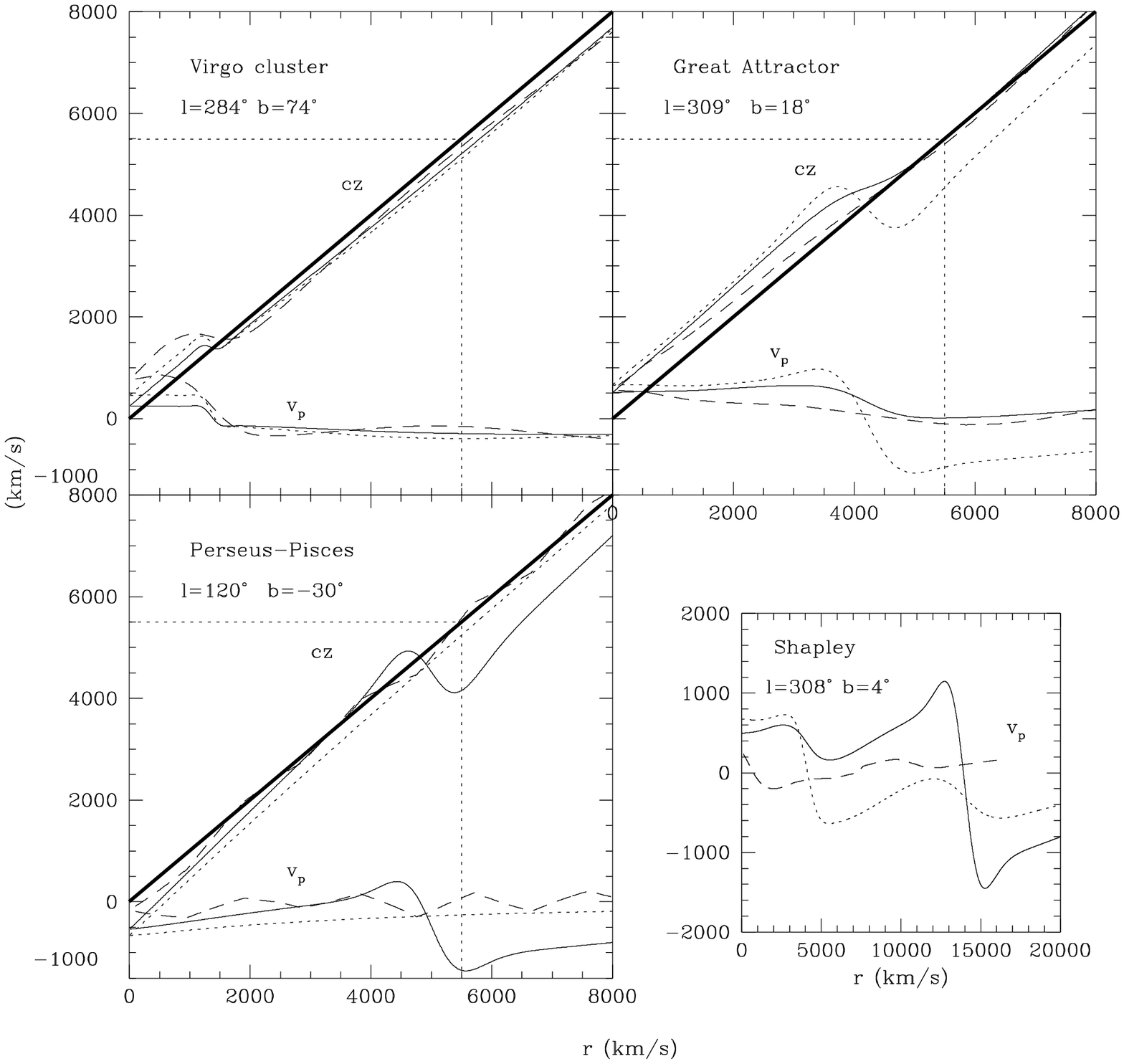}
$\ \ \ \ \ \ $\\
\vspace{15truecm}
$\ \ \ $\\
{\small\parindent=3.5mm {Fig.}~14.---
These plots present the behaviors of the recession
velocity $cz$ and peculiar velocity $v_p$ (in the CMB frame), as predicted
by some models of the peculiar velocity field, against the distance $r$
(in \ks) from the attractor center, along the lines of sight of Virgo
($l=284^{\circ}, b=74^{\circ}$), Great Attractor ($l=309^{\circ},\;
b=18^{\circ}$), and Perseus-Pisces ($l=120^{\circ},\; b=-30^{\circ}$). We
also show the behavior of the peculiar velocity along the line of sight of
the Shapley concentration ($l=308^{\circ},\; b=4^{\circ}$).  The dashed,
dotted, and solid lines refer to the predictions of the modified cluster
dipole model and multi-attractor models fitted to the Mark II subset and
the whole Mark III, respectively. The Hubble relation $cz=r$ is also
indicated. \label{velox}
}
\vspace{5mm}
\begin{multicols}{2}

Compared to Mark II, Mark III gives particularly pronounced PP and SH
attractors and a less dominant role for GA in shaping the velocity field.
SH appears to account for nearly half the LG peculiar motion. Compared to
the Mark III spirals, the velocity field of the whole Mark III, where E/S0
galaxies are included, is characterized by a greater region which is
dynamically dominated by GA, but the value of $\chi^{2}/dof$ and the
errors on the model parameters are greater. In Mark III velocity maps
there is no clear evidence of an infall pattern around PP, which rather
acts in slowing down motions towards GA in the region comprised between LG
and PP. On the other hand, large streaming flows in the general direction
of the CMB apex are particularly remarkable.

\includegraphics{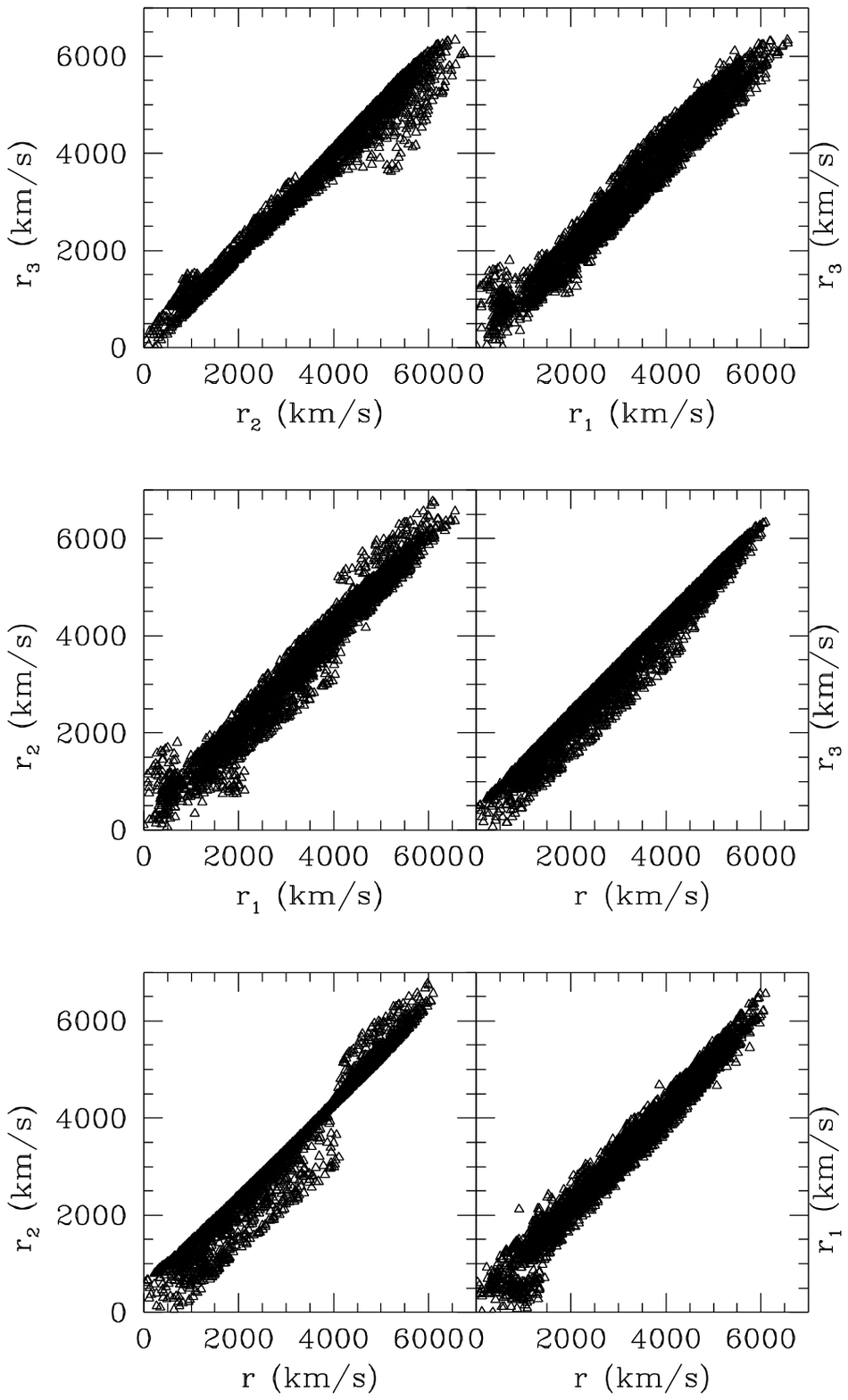}
$\ \ \ \ \ \ $\\
\vspace{13.5truecm}
$\ \ \ $\\
{\small\parindent=3.5mm {Fig.}~15.---
We show the comparisons between the uncorrected
galaxy distance $r=cz$  (expressed in \ks and in the CMB frame) 
and the three corrected galaxy distances
(hereafter $r_1$, $r_2$, $r_3$), as respectively
predicted by the modified cluster dipole model, the multi-attractor models
relative to the Mark II subsample and the Mark III spirals. \label{distconf}
}
\vspace{5mm}

As regards the cluster dipole model, the reconstructed cluster density
field (smoothed with a 1500 \ks gaussian window) appears to be in
satisfactory agreement with the Mark III POTENT density field in the
nearby region ($cz<5000$ \ks) of the supergalactic plane, where the GA
and PP superclusters appear as prominent density peaks separated by an
extended underdense region (see Plionis et al., 1996). There is less
agreement in farther regions (just where peculiar velocity data are less
reliable) so much that the Coma supercluster is replaced in Mark III
POTENT density maps by an overdense region located in a displaced
position. The corresponding velocity field maps bear less similarity than
the density field maps, because the modified cluster dipole model is
characterized by a well-defined infall pattern around PP, small large
scale flows towards GA and SH, and large Virgo infall velocities in the
local region. This large infall , which is at variance with most recent
relevant estimates (see end of \S 3.1.2), enhances the flows towards
GA in the LS region and makes the infall pattern around Coma almost 
disappearing. Thus, this large Virgocentric infall makes the modified 
cluster dipole model less dissimilar from the Mark III multi-attractor 
model in the LS region than the original cluster dipole model.

Notably, the picture which is emerging from the application of the POTENT
machinery to the new data of an ongoing peculiar velocity survey, the
I-band TF distance survey of about 2000 spiral galaxies in the field (the
SFI survey) and in the direction of 24 clusters (the SCI survey)
(Giovanelli et al., 1997 a,b), bears some features which are seen in the
modified cluster dipole model, such as a clear infall pattern around PP
and the absence of coherent streaming flows on large scale (da Costa et
al., 1996).

Differences in peculiar velocity models give rise to differences in the
predicted galaxy distances.  Fig. 15 shows the comparisons
between the uncorrected distance $r=cz$ and the three corrected distances
(hereafter $r_1$, $r_2$, $r_3$) we derive for all field galaxies and
groups of our sample, as respectively predicted by the modified cluster
dipole model, the multi-attractor models relative to the Mark II subsample
and the Mark III spirals. The $r_3$ versus $r_1$ plot reveals a poor
correlation especially at low values, where $r_3>r_1$ in many cases
(mostly because of the large Virgo infall velocity which characterizes the
first model). The $r_2$ versus $r_1$ plot reveals a poor correlation also
at high values, mostly because of strong differences in the infall towards
GA.  The $r_3$ versus $r_2$ plot shows pronounced deviations from the
one-to-one relation at low values ($r_2<2000$ \ks), where $r_2<r_3$ in
many cases (mostly because of differences in the predicted Virgocentric
infall) and at high values ($r_2>4000$ \ks), where $r_2>r_3$ for many
objects (mostly because of the smaller back-side infall towards GA given
by Mark III with respect to Mark II). But this plot displays, on average,
a smaller scatter than the two aforementioned diagrams, especially at low
and intermediate values. The comparison between corrected and uncorrected
distances shows that in general corrections are more important at low
values, as expected; however, the $r_2$ versus $r$ plot displays a large
scatter and marked systematic effects also at great values. 

\section{Conclusions}

In this paper we provide homogeneous estimates of distances for the
individual 3689 galaxies and the 485 groups (which contain a total of 2703
galaxies) of a large all-sky optical galaxy sample (Garcia et al., 1993;
Garcia, 1993), which is limited to a depth of 5500 \ks and to the 
magnitude limit of completeness of $B_T=14$ mag (for $|b|>20^{\circ}$).
This is the widest, complete, all-sky optical galaxy sample for which 
refined galaxy distances are available. All our distance estimates are 
available on request. 

We recover the distances of our objects by correcting redshift--distances
for peculiar motions through the application of some peculiar velocity
field models.

We invert the distance-redshift relations relative to different velocity
models and solve the problem of the triple-valued zones of this relation
by using blue Tully-Fisher relations calibrated on suitably defined 
samples of objects having distances predicted by peculiar velocity models. 

In our work we avoid taking inhomogeneous redshift--independent
distances obtained from various classical DIs. Homogeneous
redshift-independent distances are available for limited samples of
galaxies and it is notoriously difficult to combine together the results
coming from different DIs to achieve an uniform scale of distance.
Moreover, we do not attempt to correct redshift--distances by using the
peculiar velocity field derived from the positions and redshift of the
galaxy sample itself, because this involves delicate strategies to recover
the galaxy distribution in the unsampled zone of avoidance and because our
galaxy sample does not include all the relevant gravitational sources for
local peculiar motions. 

We regard the peculiar velocity models considered as representative of
current common views on the kinematics of cosmic flows in the nearby
universe. However, refinements of the proper calibration of the Mark III
Tully-Fisher relations, on which our velocity field models depend, are
possible. 

This point was raised by Davis, Nusser \& Willick (1996). The authors
performed a mode-by-mode comparison of the peculiar velocity fields
expressed in sets of independent basis functions, that were fitted in
redshift space to the inverse Tully--Fisher data from Mark III and to the
predictions based on the 1.2 Jy IRAS density field (Fisher et al., 1995).
Their best-fitting value of $\beta\sim$0.6 yielded no acceptable agreement
between the two velocity fields within a region of 6000 \ks radius. On the
other hand, using a new method, called VELMOD, for maximizing the
likelihood of the Tully-Fisher observables, Willick et al. (1997b) found
that the IRAS-predicted velocity field provided a satisfactory fit (with
$\beta\sim$0.5) to the Mark III Tully--Fisher data, within 3000 \ks and on
a small smoothing scale of 300 \ks, if one added an external quadrupole,
which was essentially expected from the way in which the density field was
smoothed. The discrepancy between these two approaches in comparing
velocity fields might be a result of systematic errors incurred in
matching the Mark III data subsets, an effect to which the latter approach
is insensitive.

Very recently, Willick \& Strauss (1998) applied an
implemented version of the VELMOD method to an expanded sample which
comprises nearly all Mark III non-cluster spirals to 7500 \ks. The authors
confirmed that the IRAS--predicted velocity field, with quadrupole, was
a good fit to the Tully--Fisher data of that sample, for $\beta\sim$0.5
and smoothing scales of 300 or 500 \ks. But they recognized that the
VELMOD Tully--Fisher calibration differed significantly from the Mark III
Tully--Fisher calibration for the relatively distant ($cz>$3000 \ks)
galaxies of the WCF subsample, which covers particularly the region of the
Perseus--Pisces supercluster. For these objects, the VELMOD Tully--Fisher
relations, which, however, rely on the accuracy of the IRAS--predicted
peculiar velocities, yield distances $\sim$8\% shorter than the Mark
III calibrations. The main effect is a reduction of the infall of the 
Perseus--Pisces region and, hence, of the bulk flow.   

Ongoing observations aimed at giving reliable North-South
homogenization (e.g., Strauss, 1997) will cast light on doubts
concerning the validity of the Mark III Tully--Fisher calibration.

Certainly, major developments in peculiar velocity studies will arise by
the turn of this century from analyses of peculiar velocity catalogs with
data of superior quality (in terms of sky coverage, accuracy, and
homogeneity), which can come out from peculiar velocity surveys in
progress, based on the surface brightness fluctuation (Tonry et al.,
1997), Type Ia supernovae (e.g., Riess et al., 1997), I-band Tully-Fisher
(e.g., the SCI and SFI samples by Giovanelli et al., 1997a, b), and
elliptical fundamental plane (e.g., J$\o$rgensen, Franx \& Kj$\ae$rgaard,
1996; Wegner et al., 1996; Saglia et al., 1997) methods. New corrections
to galaxy distances for peculiar motions can be easily implemented with
the aid of some techniques used in this paper. 

In any case, the use of different models of the peculiar velocity field
allows us to check to what extent differences in the current views on the
cosmic flows affect the recovering of galaxy distances in the nearby
universe. We find that differences among distance estimates appear to be
less pronounced in the $\sim$2000-4000 \ks distance range than in farther
or nearer regions (see end of \S 5). 

These differences can affect the optical luminosity function of a galaxy
sample restricted to a fairly narrow solid angle. For instance, the
addition of a Virgo infall alone of a few hundred km/s can brighten the
characteristic magnitude $M^{*}$ of the Schechter--type luminosity
function by a few tenths of magnitude , for a shallow galaxy sample (e.g.,
Efstathiou, Ellis \& Peterson, 1988). However, it can be proved
that these differences have a small effect on the
optical galaxy luminosity function 
of galaxy samples which, like our sample, cover a very large solid angle
(see Marinoni et al. 1998b for a detailed discussion). 

In general, peculiar velocity gradients are not strong so that all
galaxies of a given region share similar peculiar motions. But in many
regions of the nearby universe positive (or negative) peculiar radial
velocities of a few hundred km/s are present; in these regions our
corrections to galaxy distances draw the objects which are nearby in angular
position closer to each other (or farther from each other) by a few Mpc
with respect to their uncorrected spatial locations. Remarkably, this
effect has a minor impact on the local galaxy density on large scales
(i.e., greater than a few Mpc), whilst it affects seriously the local
galaxy density on small scales (i. e., roughly smaller than 1 Mpc) (see,
e.g., Marinoni et al., 1998a for a preliminary account). Moreover,
particularly in the above--mentioned nearest and farthest regions of the
volume considered, the latter density parameter will be particularly
sensitive to differences between the various sets of corrected distances. 

The local galaxy density on small scales is an important parameter to be
used in statistical studies of environmental effects on the properties of
nearby galaxies, since it provides a well-defined characterization of
galactic environment irrespective of membership in galaxy systems.
Remarkably, much of the observed evolution of the properties and 
populations of galaxies which has occurred during recent epochs ($z<1$) 
can be ascribed to interaction of galaxies and their local 
surroundings.

\acknowledgments

The authors are indebted to D. Burstein for his electronic distribution 
of the Mark II and Mark III datasets, A. M. Garcia for her electronic 
distribution of the data of her galaxy sample, E. Branchini 
and M. Plionis,  who provided detailed results of their computations.
 
The authors would like to thank S. Bardelli, F. Mardirossian, and M.
Mezzetti for useful conversations. 

One of the authors (P. M.) has been supported by the EC TMR grant 
ERB4001GT962279.

This work has been partially supported by the Italian Ministry of
University, Scientific and Technological Research (MURST) and by the
Italian Space Agency (ASI).

\end{multicols}
\small

\end{document}